\documentclass[journal ]{new-aiaa}
\usepackage[utf8]{inputenc}
\usepackage{textcomp}
\usepackage{graphicx}
\usepackage{amsmath}
\usepackage[version=4]{mhchem}
\usepackage{siunitx}
\usepackage{longtable,tabularx}
\usepackage[table]{xcolor}

\usepackage{subcaption}
\usepackage{multirow}
\usepackage{booktabs}
\usepackage{indentfirst}

\usepackage{comment}
\usepackage{placeins}

\usepackage{xcolor}
\hypersetup{hidelinks}

\title{Mechanism-Separated Closed-Form Transition Modeling via Field Inversion and Symbolic Regression}

\author{Seunghyun Joo\footnote{Graduate Student, Department of Aerospace Engineering, chlrh4535@snu.ac.kr}, Younghyo Kim\footnote{Postdoctoral Researcher, Department of Aerospace Engineering, hyolighting@snu.ac.kr}, and Kwanjung Yee\footnote{Professor, Department of Aerospace Engineering, kjyee@snu.ac.kr (Corresponding Author)}}
\affil{Seoul National University, Seoul 08826, Republic of Korea}

\begin{document}

\maketitle

\begin{abstract}
Next-generation aircraft, rotorcraft, and wind turbines demand improved aerodynamic efficiency, making drag reduction a central design objective; in transition-sensitive configurations, the extent of laminar flow strongly affects viscous drag and performance. Accurate prediction of laminar--turbulent transition is therefore essential. Transport-equation-based transition models, however, increase computational cost and implementation complexity. Neural-network-based closures can be difficult to interpret and integrate into independent flow solvers, while a single compact correction trained on heterogeneous transition data may fail to preserve mechanism-specific behavior. This study develops a mechanism-separated transition model using field inversion and symbolic regression. The framework treats natural, crossflow, and separation-induced transition with separate correction branches and yields explicit, closed-form corrections to the Spalart--Allmaras production term, without additional transport equations or runtime neural-network inference. The closed-form model is implemented in an independent flow solver to assess implementation portability and is evaluated on canonical cases and complex three-dimensional configurations, including a natural-laminar-flow transport wing and a hovering rotor. Across the tested cases, the model captures the principal transition-front trends and associated aerodynamic-performance changes. For the hovering-rotor case, it requires about 55\% of the wall-clock time of the comparable transport-equation transition model.
\end{abstract}

\providecommand{\keywords}[1]
{
  \textbf{\textit{Keywords---}} #1
}


\section*{Nomenclature}
{\renewcommand\arraystretch{1.0}
\noindent\begin{longtable*}{@{}l @{\quad=\quad} l@{}}

\multicolumn{2}{@{}l}{Symbols} \\
$\alpha$ & angle of attack, deg \\
$\beta$ & transition correction factor, production-term modifier \\
$\beta_{\mathrm{eff}}$ & effective transition correction factor \\
$\beta_{\mathrm{SIT},0}$ & pre-reattachment separation-branch correction \\
$\beta_{\mathrm{pre}}$ & pre-reattachment base correction \\
$\mathcal{G}_{\mathrm{SIT}}$ & separation-induced-transition gate function \\
$\mathcal{F}_{m}$ & turbulence-memory gate function \\
$\mathcal{T}_1,\ \mathcal{T}_2$ & natural-onset and turbulence-memory terms \\
$\mathrm{TC1}$ & local crossflow-transition criterion \\
$\mathcal{T}_{cf}$ & crossflow-onset term \\
$\lambda_{cf}$ & crossflow pressure-gradient parameter \\
$g_\lambda$ & crossflow pressure-gradient modulation function \\
$F_{\mathrm{reat}},\ F_{\mathrm{wake}}$ & separation-branch reattachment and wake factors \\
$F_{\mathrm{apg}},\ F_\chi$ & separation-branch adverse-pressure-gradient and turbulence-state factors \\
$C_\gamma,\ F_{\mathrm{on}}$ & reattachment-term intermittency and onset factors \\
$F_{PG}$ & pressure-gradient correction function for $Re_{\theta c}$ \\
$\mathrm{Prod.}$ & Spalart--Allmaras production term, m$^2$/s$^2$ \\
$\mathrm{Prod.}_{\mathrm{safe}}$ & safeguarded SA production term, m$^2$/s$^2$ \\
$\tilde{S}$ & Spalart--Allmaras modified vorticity, 1/s \\
$C_{b1}$ & Spalart--Allmaras production coefficient \\
$k$ & turbulent kinetic energy, m$^2$/s$^2$ \\
$Re_V$ & wall-distance vorticity Reynolds number \\
$\chi$ & eddy-viscosity ratio, $\nu_t/\nu$ \\
$C_f$ & skin-friction coefficient \\
$C_p$ & pressure coefficient \\
$C_T/\sigma$ & blade loading \\
$C_T,\ C_Q$ & rotor thrust and torque coefficients \\
$a$ & prolate-spheroid semi-major axis, m \\
$c$ & airfoil chord, m \\
$d$ & wall distance, m \\
$FM$ & figure of merit \\
$J$ & objective function for field inversion \\
$N_d,\ N_\beta,\ N_e$ & numbers of observation points, correction-field degrees of freedom, and interior-face cell pairs \\
$M$ & Mach number \\
$M_e,\ M_\infty$ & local isentropic edge and free-stream Mach numbers \\
$f_{cc}$ & compressibility correction factor for $Re_{\theta c}$ \\
$p_{0,\infty}$ & free-stream total pressure, Pa \\
$p_{\mathrm{loc}}$ & local static pressure, Pa \\
$u_\tau$ & friction velocity, m/s \\
$V$ & wall-normal velocity component, m/s \\
$i_t$ & turbulence index \\
$\nu$ & kinematic viscosity, m$^2$/s \\
$\tilde{\nu}$ & SA working variable, modified eddy viscosity, m$^2$/s \\
$\nu_t$ & turbulent eddy viscosity, m$^2$/s \\
$S,\ \Omega$ & strain-rate and rotation-rate tensors; $|\Omega|$ vorticity magnitude, 1/s \\
$\boldsymbol{\omega}$ & vorticity vector, 1/s \\
$\omega$ & reconstructed specific dissipation rate, 1/s \\
$Re$ & Reynolds number \\
$Re_{\theta}$ & momentum-thickness Reynolds number \\
$y^{+}$ & nondimensional wall distance \\
$\gamma$ & intermittency \\
$\gamma_g$ & ratio of specific heats \\
$\kappa$ & von K\'{a}rm\'{a}n constant \\
$\epsilon$ & small regularization constant \\
$\epsilon_\omega$ & vorticity-normalization regularization constant, 1/s \\
$\epsilon_P$ & production-floor constant, m$^2$/s$^2$ \\
$\theta_0$ & collective pitch angle, deg \\
$\phi$ & circumferential (azimuthal) angle, deg \\
$\lambda_{\mathrm{QoI}}$ & data-misfit weight in the field-inversion objective \\
$\lambda_{\beta}$ & regularization weight for the correction field \\
$\lambda_{\nabla}$ & spatial-smoothness regularization weight \\
$\boldsymbol{\psi}$ & adjoint variable \\

\end{longtable*}}
\setcounter{table}{0}

\section{Introduction}
\label{sec:introduction}

The development of next-generation aerodynamic systems has intensified the demand for improved aerodynamic efficiency in the aerospace and energy sectors.
As aircraft, wind-turbine, and rotorcraft designs seek to minimize energy consumption, fuel burn, and associated environmental impacts, reducing viscous drag has become a primary objective~\cite{joslin1998aircraft,shi2020natural,djeddi2024natural}. This need has motivated aerodynamic-design strategies aimed at delaying transition and preserving extended laminar-flow regions, including natural-laminar-flow airfoils and wings, laminar-flow-control concepts, and low-Reynolds-number airfoil designs intended to mitigate laminar-separation-bubble losses. In this context, accurate prediction of laminar--turbulent boundary-layer transition is essential because the transition location strongly affects skin-friction drag, heat transfer, separation behavior, and ultimately lift and moment prediction in Reynolds-averaged Navier--Stokes (RANS) simulations. Its importance has increased further in recent applications where the laminar-flow extent is no longer a secondary detail but a design objective or a dominant performance driver. For low-Reynolds-number systems such as small unmanned aerial vehicles (UAVs), micro air vehicles, eVTOL rotors, and wind-turbine blades operating at chord Reynolds numbers of approximately $Re = 10^4$--$10^6$, the transition location can strongly influence whether the boundary layer remains attached or develops a laminar separation bubble, leading to large changes in aerodynamic performance~\cite{chen2020aerodynamic, sheng2018role, furusawa2025comparative}. For high-Reynolds-number configurations, such as large transport aircraft with natural-laminar-flow (NLF) wings, accurate transition prediction is indispensable for estimating viscous drag and evaluating laminar-flow-control technologies.

Despite its profound practical importance, robust transition modeling in a RANS framework remains challenging because transition is inherently diverse and mechanism-dependent~\cite{durbin2017perspectives}. Practical aerodynamic flows may undergo natural transition, crossflow transition, or separation-induced transition, depending on the instability environment, free-stream turbulence level, pressure-gradient history, sweep, and local separation behavior. These routes to transition differ not only in the onset location but also in the physical mechanism that produces turbulence. Natural transition is associated with modal instability amplification in attached boundary layers, crossflow transition arises from three-dimensional instability mechanisms in swept flows, and separation-induced transition develops through instability growth in a separated shear layer. A transition model must therefore distinguish among these mechanisms while accounting for both local flow features and the nonlocal evolution of the boundary layer. These requirements make transition prediction particularly challenging for low-cost RANS closures. Classical transition closures have typically addressed this complexity by introducing additional transport equations and empirical correlations. A notable example is the $\gamma$--$Re_{\theta}$ transition model of Langtry and Menter~\cite{langtry2009correlation}, which has been widely used because of its compatibility with the shear-stress transport (SST) $k$--$\omega$ turbulence model and its practical formulation based on locally evaluated transition criteria~\cite{MenterLangtryVolker2006}. However, such models introduce additional transported variables and source-term closures, thereby increasing both implementation complexity and computational cost relative to a baseline one- or two-equation turbulence model. In addition, many of the empirical correlations used to trigger or localize transition are calibrated using canonical boundary-layer data, including flat-plate and related two-dimensional configurations. Extending these models to crossflow-dominated, three-dimensional, separated, or rotating flows has therefore often required mechanism-specific corrections, additional activation criteria, and case-dependent recalibration.

To circumvent the computational cost and accuracy limitations of traditional models, data-driven approaches leveraging field inversion and machine learning (FIML) have gained considerable traction~\cite{duraisamy2019turbulence}. Among these approaches, FIML, introduced by Duraisamy et al.~\cite{duraisamy2015new} and subsequently formalized by Parish and Duraisamy~\cite{parish2016paradigm}, provides a systematic framework for inferring model-form corrections from high-fidelity data. This framework has been successfully applied to turbulence production correction, eddy-viscosity augmentation, separated-flow correction~\cite{wu2025development,he2026field}, and transition modeling~\cite{yang2020improving, srivastava2022towards}. Related studies have also sought to improve model consistency and generalization by embedding machine-learning models into the inversion process or by constructing local augmentation maps in feature space~\cite{holland2019field,srivastava2021generalizable}. Nevertheless, many early FIML-based transition models rely heavily on artificial neural networks (ANNs) as the regression component~\cite{wu2022two}. Although ANNs offer strong nonlinear approximation capabilities, their black-box nature obscures the underlying physics and can hinder implementation portability because runtime inference requires coupling to external machine-learning libraries. More critically, a single global correction trained on data spanning transition mechanisms with distinct physical requirements may conflate the different correction behaviors required by those mechanisms, a failure mode termed mechanism conflation in this study and defined here as the loss of mechanism-specific correction behavior when heterogeneous transition data are represented by a single compact algebraic expression. Attached-flow natural transition, crossflow transition, and separation-induced transition require qualitatively different modifications to turbulence production and distinct activation criteria within the Spalart--Allmaras (SA) turbulence model. As demonstrated by the mixed-mechanism symbolic-regression diagnostic in Appendix~\ref{app:single_correction}, the single-expression model considered in this study exhibits an error plateau on the Pareto front despite increasing expression complexity, indicating that the tested compact global SR expressions do not represent the combined dataset with the accuracy achieved by the mechanism-specific branches under the selected features, operator library, search protocol, and complexity convention. This observation motivates a mechanism-separated formulation in which each transition route is assigned a dedicated correction branch and an admissibility gate. This strategy is conceptually consistent with the physics-based zonal modeling approach of He et al.~\cite{he2026field}, who used dedicated expressions for distinct local error modes in flows involving competing mechanisms.

Building on these observations, this study aims to address three key limitations of existing data-driven transition models: the lack of explicit formulations, limited implementation portability, and mechanism conflation. To obtain an explicit formulation that can be implemented in an independent CFD solver, symbolic regression (SR) is employed in place of ANNs~\cite{schmelzer2020discovery,tang2023discovering,wu2023enhancing}. Unlike ANNs, SR searches the space of mathematical expressions to derive explicit, closed-form corrections that can be directly inspected and embedded in computational fluid dynamics (CFD) solvers~\cite{he2024field}. To address mechanism conflation, a gated, mechanism-separated architecture is subsequently introduced. In this work, we systematically isolate three primary transition routes: natural transition, crossflow transition, and separation-induced transition. These mechanisms are selected because they are central to the low-disturbance attached-flow, swept three-dimensional, and separated-flow regimes considered in this study. Other transition mechanisms, including Görtler instability, bypass transition, roughness-induced transition, and shock-induced transition, remain outside the present scope. The same field-inversion and symbolic-regression workflow can be extended to additional mechanisms in future work, provided that suitable mechanism-specific training data, features, and gate definitions are introduced.

The mechanism-separated algebraic transition framework is constructed through a two-step data-driven procedure. Spatially varying correction fields are initially inferred via discrete-adjoint field inversion using DAFoam~\cite{he2020dafoam, he2018aerodynamic}, which optimizes the correction field using a discrete adjoint method to match reference quantities of interest while regularizing deviations toward the baseline model~\cite{wachter2006implementation, tikhonov1963regularization}. Closed-form branch expressions are subsequently obtained using PySR symbolic regression~\cite{cranmer2023interpretable} to map local flow features~\cite{pope1975more,menter2006correlation,SULUKSNA200848} to the inverted fields. Physics-informed admissibility gates restrict the crossflow and separation branches to their relevant flow regimes, while a smooth turbulence-memory gate confines the shared memory term to physically triggered regions and recovers the baseline SA behavior in fully turbulent regions~\cite{spalart1992one,rumsey2022search,bin2024constrained}. The predictive capability and cross-configuration performance of the proposed model are evaluated across diverse canonical validation cases as well as complex, mixed-mechanism three-dimensional configurations, including the CRM-NLF aircraft wing and the PSP rotor. Because the proposed correction is algebraic and embedded directly in the SA framework, it requires neither additional transition transport equations nor external machine-learning libraries at runtime. To assess implementation portability in an independent solver, the proposed formulation is implemented in KFLOW~\cite{kim2009euler,kflow2}, a CFD solver independent of the DAFoam environment used for model development. For the PSP rotor hover case, the implemented model reaches the near-asymptotic density-residual level of $L_2\approx1.5\times10^{-3}$ in approximately $2.3\times10^{3}$~s of wall-clock time. This represents an approximately 64\% wall-clock-time overhead relative to the baseline SA model, which reaches the same residual level in approximately $1.4\times10^{3}$~s. Nevertheless, the present model requires only about 55\% of the wall-clock time required by the SST Langtry--Menter transition model, which reaches the same residual level in approximately $4.2\times10^{3}$~s. These results demonstrate that the mechanism-separated FISR approach offers a practical transition-prediction strategy that can be implemented in an independent CFD solver for complex aerodynamic applications.

The remainder of this paper is organized as follows. Section~\ref{sec:modeling_methodology} presents the SA correction framework, field-inversion formulation, symbolic-regression procedure, and mechanism-separated architecture. Section~\ref{sec:training_cases} describes the training data, the construction of the mechanism-specific correction branches, and the definition of the associated gates. Section~\ref{sec:model_validation} validates the model on canonical transition configurations, and Section~\ref{sec:complex_3d_applications} assesses complex three-dimensional CRM-NLF and PSP rotor cases, focusing on physical accuracy, computational efficiency, and implementation portability. Section~\ref{sec:conclusions} summarizes the findings. Appendix~\ref{app:single_correction} presents the mixed-mechanism symbolic-regression diagnostic that motivates mechanism separation, and Appendix~\ref{app:complete_formulation} provides the complete implementation-ready model specification with the provenance and calibration of all components. The Supplemental Material contains the candidate-variable correlation and importance analyses, representative raw symbolic-regression expressions and normalization constants, the detailed sample-filtering and correction-balancing procedure, the coefficient-calibration procedure, and additional notes on the reattachment term and the component-influence analysis.

\section{Modeling methodology}
\label{sec:modeling_methodology}

\subsection{SA Production Correction Framework}
\label{sec:sa_correction_framework}

The present transition model is constructed by introducing a spatially varying scalar correction factor into the production term of the Spalart--Allmaras (SA) turbulence model. This correction locally modulates the production of the SA working variable while leaving the destruction and diffusion terms unchanged. The governing equation for the modified SA model reads

\begin{equation}
\frac{\partial \tilde{\nu}}{\partial t}
+ u_j \frac{\partial \tilde{\nu}}{\partial x_j}
=
\beta_{\mathrm{eff}} \cdot \mathrm{Prod.}
-
\mathrm{Dest.}
+
\mathrm{Diff.}
\label{eq:modified_SA}
\end{equation}

where $\tilde{\nu}$ is the SA working variable, and $\mathrm{Prod.}$, $\mathrm{Dest.}$, and $\mathrm{Diff.}$ denote the SA production, destruction, and diffusion contributions, respectively. The effective correction factor $\beta_{\mathrm{eff}}$ is a scalar field that modulates the turbulence production rate. When $\beta_{\mathrm{eff}} < 1$, production of $\tilde{\nu}$ is locally suppressed relative to the baseline SA model; when $\beta_{\mathrm{eff}} = 1$, the baseline SA production term is recovered; and when $\beta_{\mathrm{eff}} > 1$, production is locally amplified. The latter case is particularly relevant to separated transitional shear layers, where the baseline SA model may underpredict the rapid growth of turbulent viscosity required for reattachment.
Once $\beta_{\mathrm{eff}}$ is expressed as a function of local flow variables, no additional transition transport equation is required, and the modification is confined to the SA production term.

The baseline turbulence model is the Spalart--Allmaras model without the $f_{t2}$ term (the SA-noft2 variant~\cite{spalart1992one,aupoix2003extensions}), so that transition-onset control is assigned to the production correction $\beta_{\mathrm{eff}}$ rather than to the original $f_{t2}$ trip function. The free-stream value of the working variable is set to $\tilde{\nu}_\infty = 0.02\,\nu$, corresponding to a negligibly small turbulent eddy-viscosity ratio ($\nu_t/\nu \approx 0$), so that the boundary layer remains laminar upstream of the predicted transition onset.

\begin{figure}[hbt!]
\centering
\includegraphics[width=0.85\textwidth]{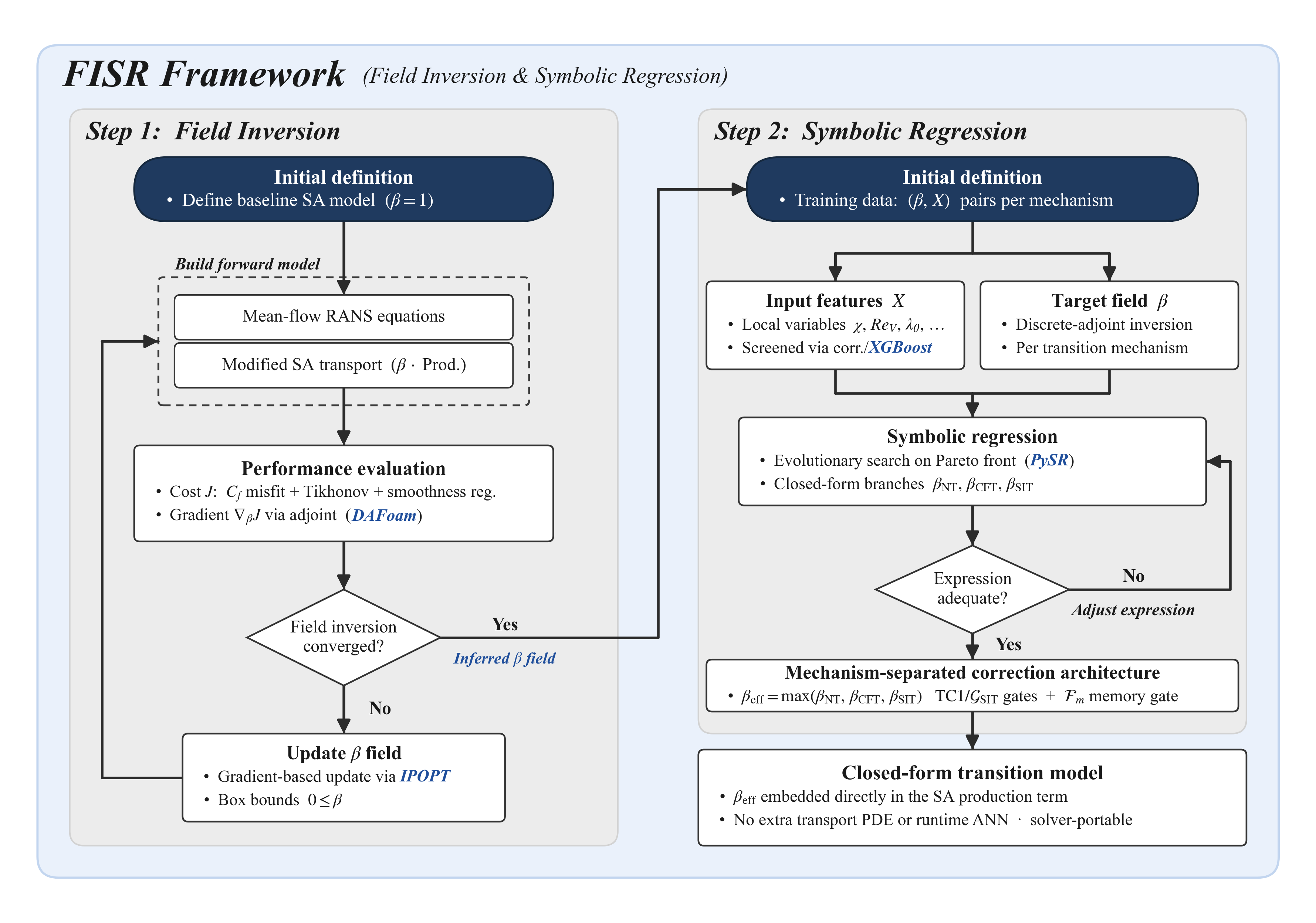}
\caption{Overview of the FISR framework for constructing the SA production correction.}
\label{fig:fisr_framework}
\end{figure}

The complete field-inversion and symbolic-regression (FISR) procedure used to obtain the algebraic production multiplier is summarized in Fig.~\ref{fig:fisr_framework}, which is organized into two main stages: Step 1 (Field Inversion) and Step 2 (Symbolic Regression). The field-inversion stage (Step 1) starts from the baseline SA model, $\beta=1$, and updates a spatial correction field $\beta(\mathbf{x})$ through repeated forward RANS--SA solves and discrete-adjoint gradients to match the reference quantity-of-interest data while regularizing the correction toward the baseline model. The converged correction fields are then paired with local flow-feature vectors to initiate Step 2. After feature reduction, symbolic regression is performed on mechanism-specific datasets, and the resulting closed-form branches are assembled with physics-informed admissibility gates and a turbulence-memory gate. In this notation, $\beta$ denotes the field-inverted correction used during training, whereas $\beta_{\mathrm{eff}}$ denotes the final mechanism-separated correction embedded in the SA production term during prediction. The following subsections describe the field-inversion formulation, flow-feature selection, symbolic-regression procedure, and mechanism-separated correction architecture in the order shown in Fig.~\ref{fig:fisr_framework}.

\subsection{Field Inversion via Discrete Adjoint Optimization}
\label{sec:field_inversion}

Given reference data from experiments, large-eddy simulations (LES), or previously validated CFD computations, the field-inversion step seeks the optimal correction field $\beta(\mathbf{x})$ that minimizes the discrepancy between the $\beta$-modified Reynolds-averaged Navier--Stokes (RANS) solution and the reference data. The problem is formulated as a constrained optimization problem:

\begin{equation}
J
=
\lambda_{\mathrm{QoI}}
\underbrace{
\sum_{i=1}^{N_d}
\left[
d_i - h_i(\beta)
\right]^2
}_{\text{QoI misfit term}}
+
\lambda_{\beta}
\underbrace{
\frac{1}{N_\beta}
\sum_{j=1}^{N_\beta}
\left(
\beta_j - 1
\right)^2
}_{\text{regularization term}}
+
\lambda_{\nabla}
\underbrace{
\frac{1}{N_e}
\sum_{(j,k)\in\mathcal{E}}
\left(
\beta_j - \beta_k
\right)^2
}_{\text{smoothness term}}
\label{eq:objective_function}
\end{equation}

where $d_i$ denotes the reference data value at observation point $i$ and $h_i(\beta)$ the corresponding RANS-predicted quantity under the correction field $\beta$; the data-misfit summation runs over the $N_d$ observation points, the Tikhonov summation runs over the $N_\beta$ correction-field degrees of freedom, and the smoothness summation runs over the set $\mathcal{E}$ of adjacent-cell pairs sharing an interior face, with $N_e=|\mathcal{E}|$. Each regularization summation is normalized by its respective count to reduce its leading dependence on the mesh size, whereas the data misfit accumulates the squared residuals over the $N_d$ wall observations, matching the discrete implementation. The smoothness term is written directly in its discrete implementation form as squared differences of $\beta$ between adjacent cells; it is dimensionless, so $\lambda_{\nabla}$ carries no length units and requires no coordinate nondimensionalization. The coefficient $\lambda_{\mathrm{QoI}}$ controls the weight of the data-misfit term, whereas $\lambda_{\beta}$ and $\lambda_{\nabla}$ control the strengths of the Tikhonov and spatial-smoothness regularization, respectively. The first term in Eq.~\eqref{eq:objective_function} measures the discrepancy between the modified RANS prediction and the reference data in terms of the selected quantity of interest (QoI), which is taken here as the wall skin-friction coefficient $C_f$. The second term is a Tikhonov regularization~\cite{tikhonov1963regularization} that penalizes deviations of $\beta$ from unity, corresponding to the baseline model. The third term is a first-order spatial-smoothness penalty on differences of the correction between adjacent cells, which discourages spatially oscillatory corrections that could reproduce similar wall observations while lying in the null space of the sparse surface data. These regularization terms serve two purposes. First, they mitigate the ill-posedness of the inverse problem, since multiple $\beta$ fields may be consistent with limited observations. Second, they bias the solution toward a smooth, baseline-consistent correction and discourage large localized corrections that are not sufficiently supported by the available data.

This particular choice of $C_f$ as the QoI is motivated by the fact that it directly reflects key transition characteristics, including the transition-onset location, the streamwise extent of transition, and the presence of laminar separation bubbles. In attached boundary-layer transition, the onset location is indicated by the characteristic rise of $C_f$ from laminar to turbulent levels. In separated transitional flows, laminar separation and turbulent reattachment are typically reflected by a near-zero or negative-$C_f$ region followed by a rapid recovery of skin friction. These characteristics make $C_f$ an informative observable for inferring the underlying transition correction field.

The optimization problem in Eq.~\eqref{eq:objective_function} is subject to the discrete governing equations of the $\beta$-modified RANS--SA system,
\begin{equation}
\mathbf{R}\left(\mathbf{Q},\boldsymbol{\beta}\right)=\mathbf{0},
\label{eq:RANS_constraint}
\end{equation}
where the state vector $\mathbf{Q}$ contains the mean-flow variables and the SA working variable at all control volumes, $\boldsymbol{\beta}$ denotes the discrete correction-field vector used as the design variable, and $\mathbf{R}$ contains the residuals of the flow equations and the modified SA transport equation. A converged forward solution defines $\mathbf{Q}$ as an implicit function of $\boldsymbol{\beta}$, so that $J=J(\mathbf{Q}(\boldsymbol{\beta}),\boldsymbol{\beta})$. Because the number of correction-field degrees of freedom is comparable to the number of grid cells, $\mathcal{O}(10^4\text{--}10^5)$ in the present work, forming the state-sensitivity matrix $d\mathbf{Q}/d\boldsymbol{\beta}$ explicitly is impractical; the discrete adjoint method eliminates it by introducing the adjoint vector $\boldsymbol{\psi}$ as the solution of the linear adjoint system
\begin{equation}
\left(
\frac{\partial \mathbf{R}}{\partial \mathbf{Q}}
\right)^T
\boldsymbol{\psi}
=
-
\left(
\frac{\partial J}{\partial \mathbf{Q}}
\right)^T ,
\label{eq:adjoint_system}
\end{equation}
after which the total gradient follows as
\begin{equation}
\nabla_{\boldsymbol{\beta}} J
=
\frac{\partial J}{\partial \boldsymbol{\beta}}
+
\boldsymbol{\psi}^{T}
\frac{\partial \mathbf{R}}{\partial \boldsymbol{\beta}} .
\label{eq:total_gradient}
\end{equation}
For each gradient evaluation, the forward problem of Eq.~\eqref{eq:RANS_constraint} is solved to convergence and the objective of Eq.~\eqref{eq:objective_function} is evaluated; the partial derivatives $\partial \mathbf{R}/\partial \mathbf{Q}$, $\partial \mathbf{R}/\partial \boldsymbol{\beta}$, $\partial J/\partial \mathbf{Q}$, and $\partial J/\partial \boldsymbol{\beta}$ are assembled consistently with the discretized residuals; the adjoint system of Eq.~\eqref{eq:adjoint_system} is solved for $\boldsymbol{\psi}$; and the gradient of Eq.~\eqref{eq:total_gradient} is passed to the nonlinear optimizer. A single adjoint solve thus yields the complete gradient with respect to all correction-field degrees of freedom for the scalar objective, which is substantially more efficient than finite-difference or direct-sensitivity approaches and makes adjoint-based gradient evaluation essential for field inversion.

Field inversion is performed using the DAFoam framework~\cite{he2020dafoam, he2018aerodynamic}, an open-source discrete-adjoint solver built upon OpenFOAM. The forward residuals are discretized using the second-order finite-volume formulation employed in the baseline DAFoam computations. The bound-constrained optimization problem is solved using IPOPT~\cite{wachter2006implementation}. 

The optimization is initialized with $\beta(\mathbf{x})=1$, corresponding to the baseline SA model, after the baseline mean-flow solution is fully converged to provide a well-defined initial state. To prevent the optimizer from exploiting nonphysical correction fields, the correction factor is constrained to be nonnegative, $\beta\geq 0$, at each correction-field degree of freedom. This constraint prevents negative production scaling; large positive corrections are discouraged by the Tikhonov regularization term. The optimization is terminated when the objective function no longer changes appreciably, assessed using the relative change in the objective function between consecutive iterations:
\begin{equation}
\frac{\left|J^{(k)}-J^{(k-1)}\right|}{\left|J^{(k-1)}\right|} < 0.01,
\label{eq:fi_convergence}
\end{equation}
where $J^{(k)}$ denotes the objective-function value at the $k$th iteration. Because each training case is inverted independently, only the within-case balance of the three terms in Eq.~\eqref{eq:objective_function} affects the inferred correction. The same nominal coefficients, $\lambda_{\mathrm{QoI}}=1$ and $\lambda_{\beta}=\lambda_{\nabla}=10^{-2}$, are used for all cases without case-specific adjustment. Because the data-misfit term is accumulated over the available observations, this choice does not imply an identical effective regularization ratio across cases. Each training case is inverted independently; the per-case problem sizes and the corresponding objective histories are presented with the training configurations in Section~\ref{sec:training_cases}.

\subsection{Local Flow Variables}
\label{sec:flow_features}

To map the RANS solution to the inferred correction field $\beta$, a set of candidate local flow variables is considered, summarized in Table~\ref{tab:input_features}. These candidates are chosen to be readily available in general CFD solvers and physically relevant to the three target transition routes, and are constructed solely from local scalar quantities and gradients of the mean-flow and turbulence fields, deliberately avoiding non-local boundary-layer integral quantities such as an explicitly integrated momentum thickness. Their information content and redundancy are examined through the correlation and feature-importance analyses, whereas their cross-configuration transferability is assessed through the validation cases.

The candidates span several physical categories. The turbulence-state group (the SA working-variable ratio $\tilde{\nu}/\nu$, the eddy-viscosity ratio $\chi=\nu_t/\nu$, and the production--destruction balance) carries the local turbulence level; because $\nu_t$ is transported by the SA equation, $\chi$ also acts as the ``memory'' of upstream transition. A group of kinematic strain--rotation invariants (the rotation-to-strain ratio, the strain- and rotation-rate invariants, the $Q$-criterion-like balance, the swirling-strength fraction, and the vortex-stretching indicator) characterizes the local velocity-gradient topology. The onset group consists of the wall-distance vorticity Reynolds number $Re_V$ and its pressure-gradient-corrected critical value $Re_{\theta c}$, which together govern natural-transition onset. The Menter pressure-gradient parameter $\lambda_\theta$ characterizes the local acceleration or deceleration state through a Galilean-invariant velocity-gradient quantity, allowing a purely local model to account for streamwise pressure-gradient effects without an additional transport equation. The crossflow measure is the vorticity-direction twist $\Psi$, which encodes the three-dimensional skewing used to detect crossflow-admissible regions. The free-stream turbulence intensity $Tu$ is prescribed as an external condition, whereas the local turbulence intensity $Tu_L$ is reconstructed from the local SA turbulence state and bounded below by $Tu$.

\begin{table}[hbt!]
\centering
\captionsetup{labelsep=period}
\caption{Candidate local flow variables considered for the mechanism-separated correction.}
\label{tab:input_features}

\begin{tabularx}{\textwidth}{@{}llX@{}}
\toprule
Variable & Expression & Physical interpretation \\ \midrule
$\tilde{\nu}/\nu$ & --- & SA working-variable ratio \\
$\chi$ & $\nu_t/\nu$ & Eddy-viscosity ratio (turbulence state and memory) \\
$P/(P{+}D)$ & $\mathrm{Prod.}/(\mathrm{Prod.}+\mathrm{Dest.})$ & SA production--destruction balance \\
$g_{\tilde{\nu}}$ & $|\nabla\tilde{\nu}|\,d/(\nu+\tilde{\nu})$ & Normalized gradient of the SA working variable \\
$Re_V$ & $|\Omega|\,d^{2}/\nu$ & Wall-distance vorticity Reynolds number~\cite{menter2006correlation} \\
$Re_{\theta c}$ & $\max(100+1000\,e^{-Tu_L F_{PG}},\,10^{-3})$ & Pressure-gradient-corrected critical $Re_\theta$~\cite{SULUKSNA200848} \\
$\lambda_\theta$ & $-7.57\times10^{-3}(\mathrm{d}V/\mathrm{d}y)\,d^{2}/\nu+0.0128$ & Menter local pressure-gradient parameter~\cite{menter2015one} \\
$|\Omega|/|S|$ & --- & Rotation-to-strain ratio \\
$\mathrm{tr}(S^2)$ & --- & Strain-rate invariant~\cite{pope1975more} \\
$\mathrm{tr}(\Omega^2)$ & --- & Rotation-rate invariant~\cite{pope1975more} \\
$\mathrm{tr}(\Omega^2 S^2)$ & --- & Rotation--strain invariant~\cite{pope1975more} \\
$Q^{*}$ & $(|\Omega|^2-|S|^2)/(|\Omega|^2+|S|^2+\epsilon)$ & Normalized $Q$-criterion balance \\
$|\Omega|/(|\Omega|+|S|)$ & --- & Swirling-strength fraction \\
$\boldsymbol{\omega}\cdot(S\boldsymbol{\omega})/|S|$ & --- & Vortex-stretching indicator \\
$\mathrm{tr}(S^3)$ & --- & Third invariant of the strain-rate tensor~\cite{pope1975more} \\
$\mathrm{tr}(\Omega^2 S)$ & --- & Rotation--strain interaction~\cite{pope1975more} \\
$\Psi$ & $|\hat{\mathbf{n}}\cdot\nabla\mathbf{e}_\omega|\,d$ & Vorticity-direction twist~\cite{grabe2013correlation} \\
$Tu$ & --- & Free-stream turbulence intensity \\
$Tu_L$ & --- & Local turbulence intensity \\
\bottomrule
\end{tabularx}

\vspace{2pt}
{\footnotesize The complete definitions of the auxiliary quantities $F_{PG}$, $Tu_L$, $\hat{\mathbf{n}}$, and $\mathbf{e}_\omega$ appearing in the expressions above are provided in Section~\ref{sec:branches}.}
\end{table}

While data-driven field inversion frameworks have been applied to general three-dimensional turbulent flows, their extension to complex three-dimensional transition mechanisms remains largely unexplored. To capture these inherently three-dimensional transitional phenomena, particular attention is devoted to the crossflow variables. Rather than computing a nonlocal crossflow Reynolds number or performing an $N$-factor integration, the model uses the local vorticity-direction twist $\Psi=|\hat{\mathbf{n}}\cdot\nabla\mathbf{e}_\omega|\,d$, a local measure inspired by the three-dimensional crossflow formulation of Grabe and Krumbein~\cite{grabe2013correlation}, where $\mathbf{e}_\omega$ is the unit vorticity vector and $\hat{\mathbf{n}}$ is the wall-normal direction. Because the vorticity direction is spatially uniform in an ideal two-dimensional boundary layer, $\Psi$ is zero there and remains small in nominally two-dimensional computations, so the crossflow-specific onset contribution is automatically suppressed, activating only where three-dimensional boundary-layer skewing develops, as in swept and rotating flows.

The information content of these candidate variables for the inferred correction is examined through a correlation and importance analysis of the field-inversion samples, reported in Supplemental Material S1, which supports the local, mechanism-specific quantities entering the branch definitions of Section~\ref{sec:branches}.

\subsection{Symbolic Regression}
\label{sec:symbolic_regression}
Symbolic regression is used to identify recurring functional motifs and to guide the selection of free coefficients within a physics-informed algebraic template, yielding a compact, closed-form algebraic representation of the inferred correction; the final branches are obtained by consolidating the recurrent SR structures with established transition indicators and bounded solver-compatible forms. Given the local variables of Table~\ref{tab:input_features} evaluated at the CFD grid cells and the field-inverted correction vector $\boldsymbol{\beta}\in\mathbb{R}^{N}$ obtained at the same locations, the objective is an explicit expression $f$ with $f(\mathbf{X})\approx\boldsymbol{\beta}$ that can be readily interpreted and embedded directly in a CFD solver. This contrasts with artificial-neural-network closures, whose internal mappings are less directly interpretable and which require coupling the solver to external machine-learning libraries at runtime.

As an initial diagnostic, a single global expression is fitted to the combined samples from natural, crossflow, and separation-induced transition. As shown in Appendix~\ref{app:single_correction}, single-expression searches in both the reduced and the full branch-feature spaces exhibit a persistent error plateau and do not attain the accuracy of the deployed mechanism-specific branches, which motivates the mechanism-separated construction. Guided by this diagnostic and by the feature-importance analysis (Supplemental Material S1), the correction is decomposed into three mechanism-specific branches. Established local transition correlations (an algebraic Bas--Cakmakcioglu natural-transition form~\cite{cakmakcioglu2020revised,mura2020revised}, an Arnal--Menter--Smirnov crossflow criterion~\cite{arnal1984theorie,mentersmirnov2014crossflow}, and a Langtry--Menter separation sensor~\cite{MenterLangtryVolker2006}) are retained as physics-informed local indicators rather than treated as discoveries of the present regression. Symbolic regression then identifies recurring mechanism-specific dependencies and guides the construction and calibration of bounded algebraic mappings from these indicators and the local turbulence state to the field-inverted correction, including the onset clipping, bounded limiting functions, and admissibility gates that appear in the resulting expressions; the associated thresholds and coefficients are reported in Section~\ref{sec:branches}. The development is therefore an FISR-assisted, physics-guided construction of a mechanism-separated closure rather than an automatic discovery of the final expressions by symbolic regression alone. The roles of literature inheritance, symbolic regression, author consolidation, and coefficient calibration for each model component are summarized in Section~\ref{sec:branches} (Table~\ref{tab:component_roles}), where the explicit expressions and coefficients are also defined.

The symbolic-regression search is performed with the PySR library~\cite{cranmer2023interpretable} using an evolutionary search over the binary operators $\{+,-,\times,\div,\min,\max\}$ and the power operator, together with the unary operators $\{\exp,\ \tanh,\ \sqrt{|\cdot|},\ \mathrm{ReLU}\}$, with a Pareto complexity penalty balancing mean-squared error against expression size. Because the operator set includes $\min$, $\max$, and $\mathrm{ReLU}$, onset clipping, bounded limiting functions, and admissibility gates can arise directly from the search rather than being appended afterward. It is used both for the mixed-mechanism diagnostic of Appendix~\ref{app:single_correction} and, in Section~\ref{sec:branches}, to identify the mechanism-specific couplings, clipping and limiting operations, admissibility gates, and free coefficients that are not fixed by the literature-derived indicators and scaling constants. The inherited correlations and fixed scales are excluded from the search and retained in their cited forms. Because the resulting corrections are explicit and algebraic, they are embedded directly into the CFD solver source code, requiring only low-cost local algebraic evaluations at runtime without any additional transport equation or case-dependent tuning.

The symbolic-regression training data are prepared as follows. The candidate local variables and the field-inverted correction $\beta$ are extracted at the transition-relevant near-wall cells ($d<0.15c$ and $Re_V>1$). To prevent the laminar ($\beta\!\approx\!0$) and fully turbulent ($\beta\!\approx\!1$) states from dominating the fit, the cells are downsampled using prescribed correction intervals with approximately equalized sampling counts (Supplemental Material S2), yielding an approximately uniform distribution over the field-inverted correction. The evolutionary search uses $40$ subpopulations of $50$ expressions with a mean-squared-error loss, and expression complexity is measured as the number of nodes in the expression tree. At each complexity level the lowest-loss expression is retained, and the reported model is selected using the PySR score, which identifies the Pareto knee where the loss reduction per unit increase in complexity is largest; the resulting complexity--MSE Pareto fronts are shown in Fig.~\ref{fig:single_sr_pareto}. For the mixed-mechanism diagnostic, the search is granted a complexity budget matched to the common-subexpression-eliminated implementation count of the complete mechanism-separated correction, following the protocol described in Appendix~\ref{app:single_correction}.

\subsection{Proposed Mechanism-Separated Correction Architecture}
\label{sec:hierarchical_formulation}

To overcome the mechanism conflation problem associated with a single universal mapping, the proposed transition model introduces a mechanism-separated correction architecture. The effective production correction is constructed through three functional pillars:

\begin{enumerate}
    \item \textbf{Mechanism-Specific Candidate Branches:} The framework evaluates three independent algebraic branches ($\beta_{\mathrm{NT}}$, $\beta_{\mathrm{CFT}}$, and $\beta_{\mathrm{SIT}}$) trained on natural, crossflow, and separation-induced transition, respectively. The overall effective correction is determined by a maximum-value branch selector:
    \begin{equation}
    \beta_{\mathrm{eff}}
    =
    \max\left(\beta_{\mathrm{NT}},\,\beta_{\mathrm{CFT}},\,\beta_{\mathrm{SIT}}\right).
    \label{eq:beta_eff}
    \end{equation}
    An inactive gated branch evaluates to zero rather than the neutral baseline multiplier ($\beta=1$). This ensures that inactive branches are excluded from the maximum selection process, preventing them from erroneously suppressing active mechanisms.

    Max and min operations are commonly used in transition closures for onset triggering, positivity preservation, and intermittency correction. For example, the Langtry--Menter $\gamma$--$Re_{\theta}$ model defines an effective intermittency of the form $\gamma_{\mathrm{eff}}=\max(\gamma,\gamma_{\mathrm{sep}})$ to account for separation-induced transition~\cite{langtry2009correlation,MenterLangtryVolker2006}. In the present model, the maximum operator selects the admissible branch that provides the largest local SA-production multiplier. Because $\beta<1$ suppresses SA production, $\beta=1$ recovers the baseline, and $\beta>1$ amplifies production, this allows the locally dominant admissible transition route to determine the correction. This choice avoids uncontrolled interactions and double counting that could arise if independently constructed branch corrections were added or multiplied.
   
    \item \textbf{Physics-Informed Admissibility:} The crossflow and separation-induced branches carry local admissibility criteria (a vorticity-direction crossflow criterion $\mathrm{TC1}$ and a vorticity-Reynolds separation gate $\mathcal{G}_{\mathrm{SIT}}$) that confine them to their physically relevant flow regimes before they enter the branch selector. Without these criteria, the crossflow branch might activate in two-dimensional attached boundary layers, or the separation-induced branch in attached high-shear regions where such corrections are physically inappropriate.

    \item \textbf{Turbulence-Memory Gate:} Because the turbulence-memory term responds to any nonzero eddy-viscosity level, it may spuriously promote transition in pre-transition regions even in the absence of a physical onset indicator. A smooth gate suppresses this contribution until one or more transition indicators become active and progressively admits it thereafter, while recovering the baseline SA behavior in fully turbulent regions.
\end{enumerate}

The attached-flow branches $\beta_{\mathrm{NT}}$ and $\beta_{\mathrm{CFT}}$ are intermittency-like algebraic production multipliers. Each is a bounded exponential-saturation multiplier constructed from a mechanism-specific local onset term (the natural-onset term $\mathcal{T}_1$ or the crossflow-onset term $\mathcal{T}_{cf}$) and a shared turbulence-memory term $\mathcal{T}_2^{\mathrm{eff}}$ that is admitted only through the smooth memory gate $\mathcal{F}_m$, following the algebraic intermittency construction of the SA-BCM transition model~\cite{cakmakcioglu2020revised,mura2020revised}; the explicit forms are given in Eqs.~\eqref{eq:beta_NT} and~\eqref{eq:beta_CFT}. Both branches remain within $[0,1]$, suppressing premature SA production upstream of transition and recovering the baseline ($\beta\rightarrow1$) once the flow becomes turbulent. The separation-induced branch $\beta_{\mathrm{SIT}}$ instead acts as a production amplifier (it may exceed unity), confined to separated shear layers by the vorticity-Reynolds gate $\mathcal{G}_{\mathrm{SIT}}$. The explicit onset terms, the memory gate, and the branch expressions, together with all inherited and calibrated coefficients, are defined in Section~\ref{sec:branches}; the maximum selector of Eq.~\eqref{eq:beta_eff} then returns the baseline value $\beta_{\mathrm{eff}}\rightarrow1$ in fully turbulent regions, where the attached-flow branches saturate to unity and $\beta_{\mathrm{SIT}}$ is switched off.

\section{Model training and construction}
\label{sec:training_cases}

\subsection{Training Datasets and Numerical Setup}

The mechanism-specific branches introduced in Section~\ref{sec:modeling_methodology} are constructed using reference datasets assembled from experiments, large-eddy simulations, and previously validated CFD computations for each target transition mechanism. Table~\ref{tab:training_cases} summarizes the training configurations, including the geometry, dominant transition mechanism, Reynolds number, free-stream turbulence intensity (FSTI), and reference-data source. Unless otherwise noted, Reynolds numbers are based on the geometric chord and free-stream conditions. For the NLF(1)-0416 and swept-wing cases, the available experiments provide the transition-onset location but not the skin-friction distribution. The inversion target is therefore taken from previously published computations whose transition onset matches the measured location, and the measured onset is used to verify that target.

\begin{table}[hbt!]
\centering
\captionsetup{labelsep=period}
\caption{Training cases and grids used for field inversion and symbolic regression ($y^{+}\le0.5$ for all cases).}
\label{tab:training_cases}

\begin{tabularx}{\textwidth}{@{}>{\raggedright\arraybackslash}p{1.9cm}>{\raggedright\arraybackslash}p{2.9cm}>{\raggedright\arraybackslash}p{2.0cm}c>{\raggedright\arraybackslash}p{2.0cm}>{\raggedright\arraybackslash}X@{}}
\toprule
Mechanism & Case & Reynolds No. & FSTI (\%) & Grid (cells) & Reference data \\ \midrule
\multirow{2}{*}{Natural} & S\&K flat plate & $3.34\times10^6$ & 0.18 & H, 150{,}000 & Exp.\ $C_f$~\cite{schubauer1956contributions} \\
 & NLF(1)-0416, $\alpha=5^\circ$ & $4\times10^6$ & 0.15 & O, 300{,}000 & Exp.\ onset~\cite{somers1981design}; CFD $C_f$~\cite{hildebrand2024grid} \\ \addlinespace
Crossflow & $45^\circ$ swept wing (NLF(2)-0415) & $2.73$, $3.73\times10^6$ & 0.09 & O, 460{,}000 & Exp.\ onset~\cite{dagenhart1989crossflow,radeztsky1993effect}; CFD $C_f$~\cite{garcia2025prediction} \\ \addlinespace
\multirow{2}{=}{Separation-induced} & NACA0012, $\alpha=4^\circ$ & $2\times10^5$ & 0.1 & O, 200{,}000 & LES $C_f$~\cite{ahmed2025data} \\
 & SD7003, $\alpha=4^\circ$ & $6\times10^4$ & 0.1 & O, 200{,}000 & LES $C_f$~\cite{ahmed2025data} \\ \bottomrule
\end{tabularx}
\end{table}

Structured computational grids are generated for all training configurations to ensure accurate boundary-layer resolution (Table~\ref{tab:training_cases}). For the airfoils and the swept-wing section, the far-field boundaries are located 50 chord lengths from the body surface, and the swept-wing simulations additionally apply periodic boundary conditions in the spanwise direction to represent an infinite-span configuration. Near-wall spacing is strictly maintained at $y^{+} \le 0.5$ across all cases, and grid-refinement checks indicate that the selected mesh densities provide sufficiently converged $C_f$ distributions for use as the quantity of interest in the field-inversion procedure. For the field-inversion step of Section~\ref{sec:modeling_methodology}, the per-case problem sizes are $N_d=500$ and $N_\beta=1.5\times10^{5}$ for the Schubauer--Klebanoff (S\&K) flat plate, $750$ and $3.0\times10^{5}$ for the NLF(1)-0416 airfoil, $920$ and $4.6\times10^{5}$ for the $45^\circ$ swept wing, and $720$ and $2.0\times10^{5}$ for each of the NACA0012 and SD7003 airfoils.

\begin{figure}[hbt!]
\centering
\includegraphics[width=0.48\textwidth]{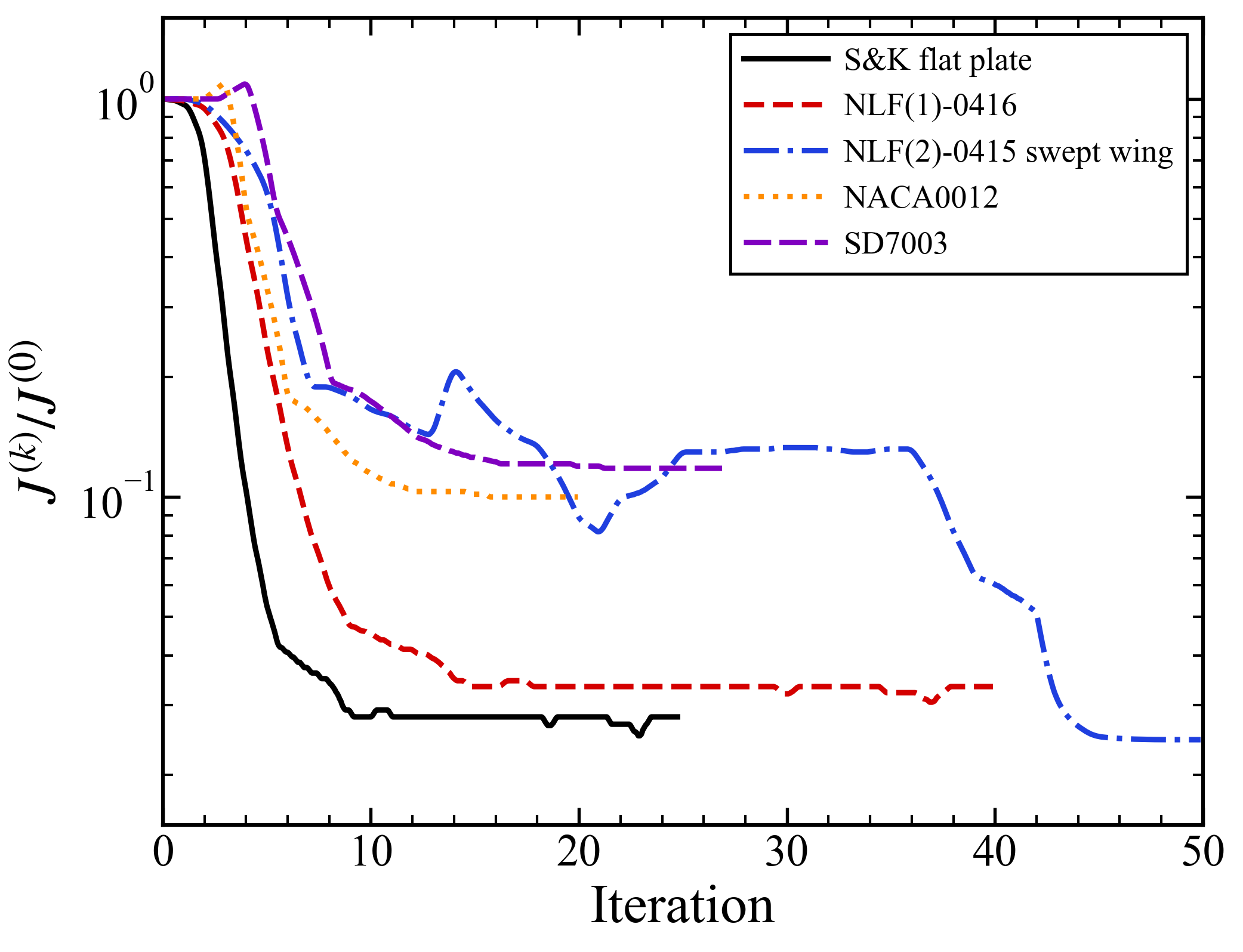}
\caption{Normalized field-inversion objective histories, $J^{(k)}/J^{(0)}$, for the five independently inverted training configurations. Each curve ends at the iteration at which the relative-change termination criterion of Eq.~\eqref{eq:fi_convergence} is met.}
\label{fig:fi_histories}
\end{figure}

Figure~\ref{fig:fi_histories} shows the objective histories of the five independently inverted training configurations, each normalized by its initial value, $J^{(k)}/J^{(0)}$. All histories stabilize under the termination criterion of Eq.~\eqref{eq:fi_convergence}. Because the inversions are performed independently and the absolute objective scales differ among the configurations, the absolute levels are not compared across cases; the normalized histories are shown only to demonstrate within-case stabilization. The histories are not required to decrease monotonically at every outer iteration, because the bound-constrained interior-point optimizer employs a filter line-search strategy with adaptive barrier-parameter updates~\cite{wachter2006implementation}, which produces the temporary increases and intermediate plateaus visible in the crossflow- and separation-case histories.

\subsection{Mechanism-Specific Branches and Gate Definitions}
\label{sec:branches}
Building on the architecture of Section~\ref{sec:hierarchical_formulation}, this subsection defines the explicit expressions and coefficients of the three branches. It first introduces the common local variables and the natural-transition threshold, then the crossflow indicators, then the attached-flow branches and the turbulence-memory gate, and finally the separation-induced branch. Each branch combines established local transition indicators with algebraic mappings, admissibility gates, clipping operations, and limiting functions identified from the corresponding field-inversion data by symbolic regression. Coefficients belonging to the underlying literature indicators are retained from the cited sources: the onset and memory scaling constants $0.002$ and $50$ are inherited from the underlying algebraic transition model~\cite{cakmakcioglu2020revised,mura2020revised}, whereas the memory-gate threshold and exponents ($0.45$, $24$, and $12$), the separation-gate coefficient of $5$, the limiter scales ($\chi=20$, $Re_V=10^{5}$, $\lambda_\theta=0.05$, and $Re_V=2420$), and the separation-branch amplitude and upper bound are identified through the mechanism-specific search. Table~\ref{tab:component_roles} summarizes, for each component, what is inherited from the literature, which recurrent motif the mechanism-specific search identified, what the authors consolidated, and which constants were subsequently calibrated; the per-coefficient provenance, search ranges, and selection criteria are detailed in Table~\ref{tab:provenance}.

\begin{table}[hbt!]
\centering
\captionsetup{labelsep=period}
\caption{Roles of literature inheritance, symbolic regression, and author consolidation in the model components.}
\label{tab:component_roles}
\begin{tabularx}{\textwidth}{@{}>{\raggedright\arraybackslash}p{2.2cm}>{\raggedright\arraybackslash}X>{\raggedright\arraybackslash}X>{\raggedright\arraybackslash}X>{\raggedright\arraybackslash}X@{}}
\toprule
Component & Inherited from literature & Recurrent SR motif & Author consolidation & Calibrated constants \\ \midrule
Natural onset & BCM onset form; $Re_{\theta c}(F_{PG})$ correlation; scales $0.002$, $50$ & Clipped onset-margin ramp in $Re_\theta-Re_{\theta c}$ & Mapped onto the exponential intermittency form & None (inherited scales fixed) \\ \addlinespace
Memory gate $\mathcal{F}_m$ & BCM memory term $50\chi$ & Turbulence-memory offset via $\chi$ suppressing re-triggering & Smooth OR-surrogate gate & Threshold $0.45$; exponents $24$, $12$ \\ \addlinespace
Crossflow branch & AMS criterion $\mathrm{TC1}$, $g_\lambda$ & Saturated ramp in $\Psi\,Re_V\,\chi$ & Folded into the exponential intermittency form & None beyond inherited \\ \addlinespace
Separation branch & Langtry--Menter separation sensor $Re_V/Re_{\theta c}$ & Bounded $\sqrt{Re_V}$ amplification gated by $\chi$ & Limiter product and reattachment term & Gate coefficient $5$; $\chi=20$; $Re_V=10^5$, $2420$; $\lambda_\theta=0.05$; amplitude $2$, bound $10$ \\ \addlinespace
Combination & --- & --- & $\max$ selector with admissibility gates & None \\ \addlinespace
$f_{cc}$ & Flight-test transition data~\cite{fisher1982inflight} & --- & Author least-squares fit & Fit coefficients \\ \bottomrule
\end{tabularx}
\end{table}

\subsubsection{Common Local Variables and Natural-Transition Threshold}

Natural transition occurs through the exponential amplification of Tollmien--Schlichting (TS) wave instabilities in low-disturbance attached boundary layers~\cite{van1956suggested}. Its onset is governed by accumulated modal amplification, commonly characterized by the $N$-factor in the $e^N$ framework. A local algebraic SA correction cannot reconstruct this modal history exactly; its role is to suppress premature SA production in attached pre-transition regions and to recover it once a local Reynolds-number threshold is exceeded. The natural branch is calibrated using the S\&K flat-plate data~\cite{schubauer1956contributions} and the NLF(1)-0416 airfoil data at $\alpha=5^\circ$, with the flat-plate case providing the primary zero-pressure-gradient reference.

The natural-onset term, defined below, compares an algebraic proxy for the local momentum-thickness Reynolds number, estimated from the wall-distance vorticity Reynolds number,
\begin{equation}
Re_\theta=\frac{Re_V}{2.193},\qquad Re_V=\frac{d^{2}|\Omega|}{\nu},
\label{eq:re_theta}
\end{equation}
with a pressure-gradient-corrected critical value $Re_{\theta c}$. The correction follows the Menter local formulation using the Galilean-invariant pressure-gradient parameter
\begin{equation}
\lambda_\theta=-7.57\times10^{-3}\,\frac{(\mathrm{d}V/\mathrm{d}y)\,d^{2}}{\nu}+0.0128,\qquad
\frac{\mathrm{d}V}{\mathrm{d}y}=\nabla(\hat{\mathbf{n}}\cdot\mathbf{U})\cdot\hat{\mathbf{n}},
\label{eq:lambda_theta}
\end{equation}
where $\hat{\mathbf{n}}=\nabla d/|\nabla d|$ is the wall-normal direction. Under the boundary-layer approximation, for which the wall-normal pressure gradient is negligible, this Galilean-invariant velocity-gradient quantity serves as a local surrogate for streamwise pressure-gradient effects without requiring a separate transport equation. The pressure-gradient function and the critical Reynolds number are
\begin{equation}
F_{PG}=
\begin{cases}
\min(1+14.68\,\lambda_\theta,\;1.5), & \lambda_\theta\ge0,\\[2pt]
\min\!\left[1-7.34\,\lambda_\theta-7.34\min(\lambda_\theta+0.0681,0),\;3.0\right], & \lambda_\theta<0,
\end{cases}
\label{eq:fpg}
\end{equation}
\begin{equation}
Re_{\theta c}=\max\!\left[100+1000\exp(-Tu_L\,F_{PG}),\;10^{-3}\right],
\label{eq:re_thetac}
\end{equation}
where the local turbulence intensity $Tu_L$ (in percent) is reconstructed from the SA eddy viscosity and floored at the free-stream value $Tu$,
\begin{equation}
\omega=\frac{|\Omega|}{\sqrt{0.09}},\qquad
k=\nu_t\,\omega,\qquad
Tu_L=\max\!\left[\min\!\left(100\,\frac{\sqrt{2k/3}}{\omega\,d},\;100\right),\;Tu\right].
\label{eq:tuL}
\end{equation}
Through $F_{PG}$, an adverse pressure gradient ($\lambda_\theta<0$) increases $F_{PG}$ and thereby lowers $Re_{\theta c}$, promoting onset, whereas in the favorable region $F_{PG}$ remains bounded ($F_{PG}\le1.5$) and $Re_{\theta c}$ stays higher, so that transition is advanced toward the adverse-gradient region downstream of the pressure minimum. For compressible applications, $Re_{\theta c}$ is additionally scaled by a compressibility factor $f_{cc}$, constructed by fitting the Mach-number dependence of the transition Reynolds number measured in the $10^\circ$-cone in-flight experiments of Fisher and Dougherty~\cite{fisher1982inflight}; the factor remains approximately unity in the incompressible limit ($f_{cc}(M_e{=}0)\approx1.008$) and is not applied in the incompressible cases,
\begin{equation}
Re_{\theta c}\;\leftarrow\;f_{cc}\,Re_{\theta c},\qquad
f_{cc}=\sqrt{\,1+\frac{3.3}{1+\exp\!\left[-6.698\left(1.2\,M_e-0.7913\right)\right]}\,},
\label{eq:fcc_main}
\end{equation}
where the local isentropic edge Mach number $M_e$ is recovered from the local static pressure $p_{\mathrm{loc}}$ and the free-stream total pressure $p_{0,\infty}$,
\begin{equation}
M_e=\sqrt{\max\!\left[0,\;\frac{2}{\gamma_g-1}\left(\left(\frac{p_{0,\infty}}{p_{\mathrm{loc}}}\right)^{\!(\gamma_g-1)/\gamma_g}\!-1\right)\right]},\qquad
p_{0,\infty}=p_{\infty}\left(1+\frac{\gamma_g-1}{2}M_{\infty}^{2}\right)^{\!\gamma_g/(\gamma_g-1)},
\label{eq:me_main}
\end{equation}
with $\gamma_g=1.4$ the ratio of specific heats and $M_{\infty}$ the free-stream Mach number; $f_{cc}$ increases $Re_{\theta c}$ with rising local Mach number, delaying transition onset in compressible regions.

With $Re_\theta$ and $Re_{\theta c}$ now defined, the natural-onset term compares the two through a clipped onset-margin ramp, and the shared turbulence-memory term is proportional to the eddy-viscosity ratio,
\begin{equation}
\mathcal{T}_1=\frac{\max\left(Re_\theta-Re_{\theta c},\,0\right)}{0.002\,Re_{\theta c}},\qquad
\mathcal{T}_2=50\,\chi=50\,\frac{\nu_t}{\nu},
\label{eq:T1}
\end{equation}
where the onset and memory scaling constants $0.002$ and $50$ are inherited from the SA-BCM model~\cite{cakmakcioglu2020revised,mura2020revised}. The natural branch $\beta_{\mathrm{NT}}$ built from $\mathcal{T}_1$ and the gated memory term is given in Eq.~\eqref{eq:beta_NT} of Section~\ref{sec:attached_branches}.

\subsubsection{Crossflow Transition Model}

Crossflow transition arises in three-dimensional boundary layers when the external pressure-gradient and local streamline directions are misaligned, producing a crossflow component within the boundary layer~\cite{saric2003stability}. In swept-wing flows, the resulting crossflow-velocity profile is inflectional and supports crossflow instability modes distinct from Tollmien--Schlichting waves~\cite{dagenhart1999crossflow}. In a purely local algebraic RANS closure, this nonlocal process is detected through the vorticity-direction twist
\begin{equation}
\Psi=\left|\hat{\mathbf{n}}\cdot\nabla\mathbf{e}_\omega\right|d,\qquad
\mathbf{e}_\omega=\frac{\boldsymbol{\omega}}{\max\!\left(|\boldsymbol{\omega}|,\,\epsilon_\omega\right)},
\label{eq:psi}
\end{equation}
a local normalized-vorticity-direction measure used in the crossflow transition criterion of Menter and Smirnov~\cite{mentersmirnov2014crossflow} and related local crossflow formulations~\cite{grabe2013correlation} ($\mathbf{e}_\omega$ is the unit vorticity vector, whose normalization is regularized by a small constant $\epsilon_\omega$, equal to $10^{-15}$~s$^{-1}$ in the implementation as stated in Appendix~\ref{app:complete_formulation}, to avoid singular behavior where the vorticity magnitude vanishes). Following the local crossflow formulation of Menter and Smirnov~\cite{mentersmirnov2014crossflow}, as subsequently implemented by Nichols~\cite{nichols2019addition} and applied within an SA-$\gamma$ framework by Lee and Baeder~\cite{leebaeder2021prediction}, a pressure-gradient-dependent modulation function $g_\lambda$ is constructed from the local crossflow pressure-gradient parameter $\lambda_{cf}$,
\begin{equation}
\lambda_{cf}=\min\!\left[\max\!\left(-7.57\times10^{-3}\,\frac{(\mathrm{d}V/\mathrm{d}y)\,d^{2}}{\nu}+0.0174,\;0\right),\;0.0477\right],
\label{eq:lambda_cf}
\end{equation}
\begin{equation}
g_\lambda=\min\!\left[\max\!\left(\left((27864\,\lambda_{cf}-1962)\lambda_{cf}+54.3\right)\lambda_{cf}+1,\;1\right),\;2.3\right],
\label{eq:cf_glambda}
\end{equation}
which enters the local crossflow criterion and the crossflow-onset term
\begin{equation}
\mathrm{TC1}=\frac{0.684}{150.8\,g_\lambda}\,\Psi\,Re_V,\qquad
\mathcal{T}_{cf}=\frac{\max(\mathrm{TC1}-1,\;0)}{0.002}.
\label{eq:tc1_tcf}
\end{equation}
The pressure-gradient variable $\lambda_{cf}$, the modulation function $g_\lambda$, the vorticity-direction twist $\Psi$, and the resulting criterion $\mathrm{TC1}$, which ultimately derives from the Arnal C1 crossflow criterion~\cite{arnal1984theorie}, are retained as literature-derived physical indicators rather than as outputs of the present symbolic regression. Using $\mathrm{TC1}$ and the selected turbulence-state variables as inputs, the mechanism-specific procedure instead identifies the crossflow-onset transformation $\mathcal{T}_{cf}$, its coupling to the shared memory gate, and the bounded crossflow branch $\beta_{\mathrm{CFT}}$. The direct crossflow-onset term $\mathcal{T}_{cf}$ becomes active when $\mathrm{TC1}$ exceeds unity, after which the crossflow branch $\beta_{\mathrm{CFT}}$, defined in the following subsection, increases in response to $\mathcal{T}_{cf}$ and the gated memory term. Because $\Psi\equiv0$ in nominally two-dimensional boundary layers, $\mathrm{TC1}\equiv0$ and $\mathcal{T}_{cf}\equiv0$ in such flows. The crossflow-specific onset contribution is therefore automatically suppressed without requiring an explicit two-dimensional mask.

\subsubsection{Attached-Flow Branches and Turbulence-Memory Gate}
\label{sec:attached_branches}

With the natural-onset term $\mathcal{T}_1$, the crossflow-onset term $\mathcal{T}_{cf}$, and the memory term $\mathcal{T}_2$ defined above, the shared memory term enters the attached-flow branches only through a smooth gate that approximates the logical union of three activation conditions: natural-transition onset, characterized by $\mathcal{T}_1$; crossflow activity, characterized by $\mathrm{TC1}$ relative to $0.45$; and an established turbulent state, characterized by $\chi$ relative to $2$,
\begin{equation}
\mathcal{T}_2^{\mathrm{eff}}=\mathcal{T}_2\,\mathcal{F}_{m},\qquad
\mathcal{F}_{m}=1-\exp\!\left(-\left[\mathcal{T}_1+\left(\frac{\mathrm{TC1}}{0.45}\right)^{24}+\left(\frac{\chi}{2}\right)^{12}\right]\right).
\label{eq:gate}
\end{equation}
The high exponents provide sharp but continuous activation of the crossflow and established-turbulence contributions, while the natural-onset contribution varies continuously with $\mathcal{T}_1$; the resulting gate approximates a logical union of the activation conditions without a discontinuous Boolean switch, although it does not produce an exact zero--one activation. Because $\mathcal{F}_{m}\rightarrow1$ when the flow is fully turbulent ($\chi\gg2$) or when a transition indicator becomes sufficiently strong, the gate prevents the memory term from independently initiating transition and restricts its role to sustaining an established turbulent state, thereby avoiding spurious transition onset in the absence of a physical trigger, in the spirit of the shielding functions used in data-enhanced transition and conditioned field-inversion models~\cite{ahmed2025data,wu2025development}. The lower crossflow reference value $\mathrm{TC1}=0.45$ enters only through this gate, allowing the turbulence-memory contribution to be introduced progressively before the direct crossflow-onset threshold $\mathrm{TC1}=1$ is reached.

The natural- and crossflow-transition branches are then the bounded exponential-saturation multipliers
\begin{align}
\beta_{\mathrm{NT}} &= 1-\exp\!\left(-\sqrt{\mathcal{T}_1}-\sqrt{\mathcal{T}_2^{\mathrm{eff}}}\right),
\label{eq:beta_NT}\\
\beta_{\mathrm{CFT}} &= 1-\exp\!\left(-\sqrt{\mathcal{T}_{cf}}-\sqrt{\mathcal{T}_2^{\mathrm{eff}}}\right),
\label{eq:beta_CFT}
\end{align}
whose form follows the algebraic intermittency construction of the SA-BCM transition model~\cite{cakmakcioglu2020revised,mura2020revised}, with the onset and memory terms redefined for each mechanism. Both branches remain within $[0,1]$: they suppress premature SA production upstream of transition and recover the baseline ($\beta\rightarrow1$) once either the onset term or the memory term grows large. Because natural transition is the default attached-flow route, the natural branch carries no separate admissibility gate; the memory gate $\mathcal{F}_m$ suppresses the memory term in the absence of a transition indicator and progressively activates it as the trigger variables increase, and the saturation of $\beta_{\mathrm{NT}}$ recovers the baseline SA behavior once the flow is turbulent.

\subsubsection{Separation-Induced Transition Model}

Separation-induced transition occurs when a laminar boundary layer separates under an adverse pressure gradient, forming an inflectional separated shear layer whose Kelvin--Helmholtz-type instability triggers transition and turbulent reattachment~\cite{tani1964low,dovgal1994laminar}. The resulting laminar separation bubble shows near-zero or negative skin friction in the separated region followed by a rapid recovery~\cite{alam2000direct}. Unlike the attached-flow branches, $\beta_{\mathrm{SIT}}$ is designed to amplify SA production, exceeding unity when the separation-induced-transition conditions are sufficiently strong; it is intended to reinforce an incipient transition in the separated shear layer rather than to initiate transition in an otherwise undisturbed laminar flow. It is calibrated on LES data for the NACA0012 ($\alpha=4^\circ$, $Re=2\times10^5$) and SD7003 ($\alpha=4^\circ$, $Re=6\times10^4$) airfoils.

The branch is built from a vorticity-Reynolds onset gate and a product of limiting functions,
\begin{equation}
\mathcal{G}_{\mathrm{SIT}}=\min\!\left[\max\!\left(\frac{Re_V/(5\,Re_{\theta c})-1}{0.5},\;0\right),\;1\right],
\label{eq:gate_sit}
\end{equation}
\begin{equation}
\beta_{\mathrm{SIT},0}=\min\!\left[2\,\mathcal{G}_{\mathrm{SIT}}\,F_{\mathrm{reat}}\,F_{\mathrm{wake}}\,F_{\mathrm{apg}}\,F_\chi,\;\;10\right],
\label{eq:beta_sit0}
\end{equation}
with the limiters
\begin{equation}
F_{\mathrm{reat}}=\exp\!\left[-\left(\frac{\chi}{20}\right)^{4}\right],\quad
F_{\mathrm{wake}}=\exp\!\left[-\min\!\left(\frac{Re_V}{10^{5}},\,30\right)^{2}\right],\quad
F_{\mathrm{apg}}=\min\!\left[\max\!\left(\frac{-\lambda_\theta}{0.05},\,0\right),\,1\right],\quad
F_\chi=\frac{\chi}{5}.
\label{eq:sit_limiters}
\end{equation}
The reattachment term is then added to $\beta_{\mathrm{SIT},0}$ using the pre-reattachment base correction $\beta_{\mathrm{pre}}$, the maximum of the intermittency-level branches evaluated before the reattachment term:
\begin{equation}
\beta_{\mathrm{pre}}=\max\!\left(\beta_{\mathrm{NT}},\,\beta_{\mathrm{CFT}},\,\beta_{\mathrm{SIT},0}\right),\qquad
C_\gamma=\min\!\left[\frac{\max(\beta_{\mathrm{pre}}-0.2,\,0)}{0.8},\,1\right],\qquad
F_{\mathrm{on}}=\min\!\left[\max\!\left(\frac{Re_V}{2420}-1,\,0\right),\,3\right],
\label{eq:sit_reat}
\end{equation}
\begin{equation}
\beta_{\mathrm{SIT}}=\beta_{\mathrm{SIT},0}
+\frac{C_\gamma\,F_{\mathrm{on}}\,\max(3\nu-\nu_t,\,0)\,|\Omega|}{\mathrm{Prod.}_{\mathrm{safe}}},\qquad
\mathrm{Prod.}_{\mathrm{safe}}=C_{b1}\tilde{S}\tilde{\nu}+\epsilon_{P},
\label{eq:beta_sit}
\end{equation}
where $\mathrm{Prod.}_{\mathrm{safe}}$ is the baseline SA production of Eq.~\eqref{eq:modified_SA} augmented with the additive floor $\epsilon_{P}=10^{-30}$~m$^2$/s$^2$ used in the implementation, which prevents division by zero where $\tilde{S}\tilde{\nu}$ vanishes while leaving the term unchanged elsewhere.
The onset gate $\mathcal{G}_{\mathrm{SIT}}$ becomes active where the wall-distance vorticity Reynolds number exceeds the separated-flow threshold $5\,Re_{\theta c}$; this vorticity-Reynolds onset gate follows the separated-flow correction of the Langtry--Menter $\gamma$--$Re_{\theta}$ model~\cite{MenterLangtryVolker2006}, and the mechanism-specific search yields the threshold coefficient of $5$, compared with $3.235$ in the original Langtry--Menter separation sensor. $F_{\mathrm{reat}}$ progressively attenuates the amplification as the eddy-viscosity ratio exceeds the characteristic reattachment scale $\chi=20$; $F_{\mathrm{wake}}$ suppresses the correction where $Re_V$ becomes excessively large; and the adverse-pressure-gradient gate $F_{\mathrm{apg}}$, which is nonzero only for $\lambda_\theta<0$, restricts the branch to adverse-gradient regions and prevents spurious activation in favorable-gradient attached layers, where the $Re_V/Re_{\theta c}$ sensor could otherwise become artificially large. The turbulence-state factor $F_\chi=\chi/5$ makes $\beta_{\mathrm{SIT}}$ proportional to the locally accumulated eddy viscosity, which is transported by the SA equation and can accumulate within the recirculation region; the branch therefore vanishes in the limit $\chi\to0$ and reinforces, rather than independently initiates, transition in the separated shear layer. The second term in $\beta_{\mathrm{SIT}}$, gated by $C_\gamma$ and $F_{\mathrm{on}}$ and normalized by the safeguarded SA production $\mathrm{Prod.}_{\mathrm{safe}}$, is a self-limiting reattachment contribution that accelerates the growth of eddy viscosity toward turbulent reattachment after the pre-reattachment base correction exceeds the prescribed activation threshold ($\beta_{\mathrm{pre}}>0.2$) while the eddy viscosity has not yet developed; it vanishes once $\nu_t$ exceeds $3\nu$ and is designed to act primarily in transitional separated regions and during reattachment. Because $C_\gamma$ depends only on $\beta_{\mathrm{pre}}$ of Eq.~\eqref{eq:sit_reat}, which is evaluated before the reattachment term is added, the correction is computed sequentially and involves no self-referential dependence.

A consolidated implementation-ready specification of the complete formulation, including all constants, clipping operations, coefficient provenance, and the evaluation order, is provided in Appendix~\ref{app:complete_formulation}.

\section{Model validation}
\label{sec:model_validation}

The final mechanism-separated correction constructed in Section~\ref{sec:training_cases} is validated using the same RANS solver employed for field inversion, but with the field-inverted correction replaced by the closed-form algebraic expressions.

The grid specifications for the validation cases are summarized in Table~\ref{tab:validation_grids}. Consistent with the training configurations, all validation grids maintain a near-wall spacing of $y^{+}<0.5$, and the far-field boundaries for the two-dimensional airfoils are placed at $50c$. 

\begin{table}[hbt!]
\centering
\captionsetup{labelsep=period}
\caption{Grid specifications for the validation cases.}
\label{tab:validation_grids}

\begin{tabular}{@{}lllll@{}}
\toprule
Mechanism & Configuration & Grid Type & Total Cells & $y^{+}$ \\ \midrule
Natural & NLF(1)-0416 & Structured O-grid & 300{,}000 & 0.5 \\ \addlinespace
Crossflow & 6:1 Prolate spheroid & Structured multi-block & 12{,}000{,}000 & 0.5 \\ \addlinespace
\multirow{2}{*}{Separation-induced} & NACA0015 & Structured O-grid & 200{,}000 & 0.5 \\
 & SD7003 & Structured O-grid & 152{,}000 & 0.5 \\ \bottomrule
\end{tabular}
\end{table}

\subsection{Natural Transition: NLF(1)-0416 Airfoil}

The natural-transition branch is first evaluated over the NLF(1)-0416 natural-laminar-flow airfoil at $Re=4\times10^6$, $M=0.1$, and $\mathrm{FSTI}=0.15\%$. Because the natural-transition branch $\beta_{\mathrm{NT}}$ was calibrated using both the S\&K flat plate and the NLF(1)-0416 airfoil at $\alpha=5^\circ$, the aerodynamic polar assesses the model at angles of attack beyond the single NLF(1)-0416 incidence included in training. The NLF(1)-0416 airfoil is designed to maintain an extended favorable-pressure-gradient region, delaying transition toward the pressure-minimum region where the pressure gradient changes from favorable to adverse. The simulation results are compared against the experimental data of Somers~\cite{somers1981design}.

\begin{figure}[hbt!]
\centering
\includegraphics[width=1.0\textwidth]{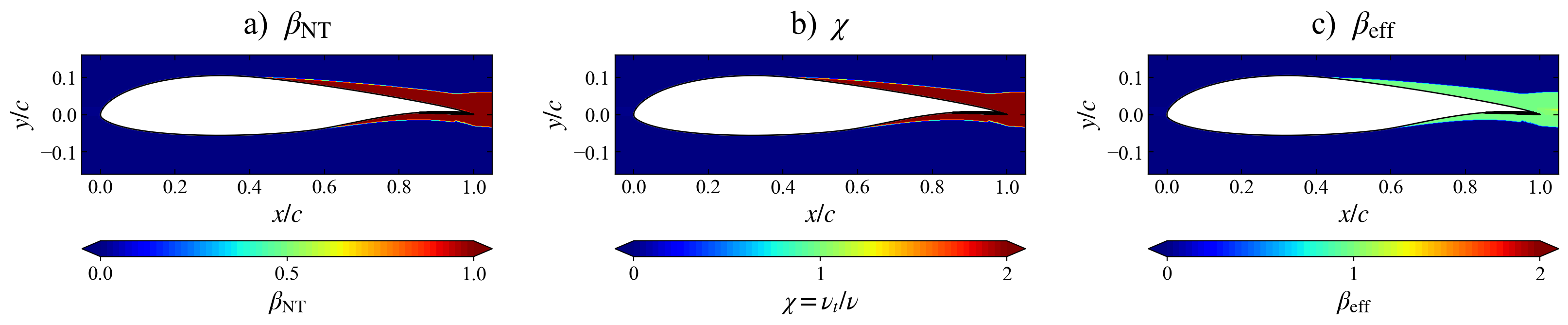}
\caption{Natural-transition branch fields from the present model on the NLF(1)-0416 airfoil ($Re=4\times10^6$, $M=0.1$, $\alpha=0^\circ$): (a) natural-branch correction $\beta_{\mathrm{NT}}$, (b) eddy-viscosity ratio $\chi=\nu_t/\nu$, and (c) effective correction $\beta_{\mathrm{eff}}$.}
\label{fig:nlf_branch_activation}
\end{figure}

Figure~\ref{fig:nlf_branch_activation} summarizes the natural-branch activity for this case. Because no separate natural-transition gate is used, the relevant diagnostic quantities are the natural-branch correction $\beta_{\mathrm{NT}}$, the effective correction $\beta_{\mathrm{eff}}$, and the eddy-viscosity-ratio field $\chi$ that governs the memory gate. The maps are intended to verify that the correction suppresses production in the attached laminar region and recovers baseline SA behavior downstream of transition.

\begin{figure}[hbt!]
\centering
\includegraphics[width=0.95\textwidth]{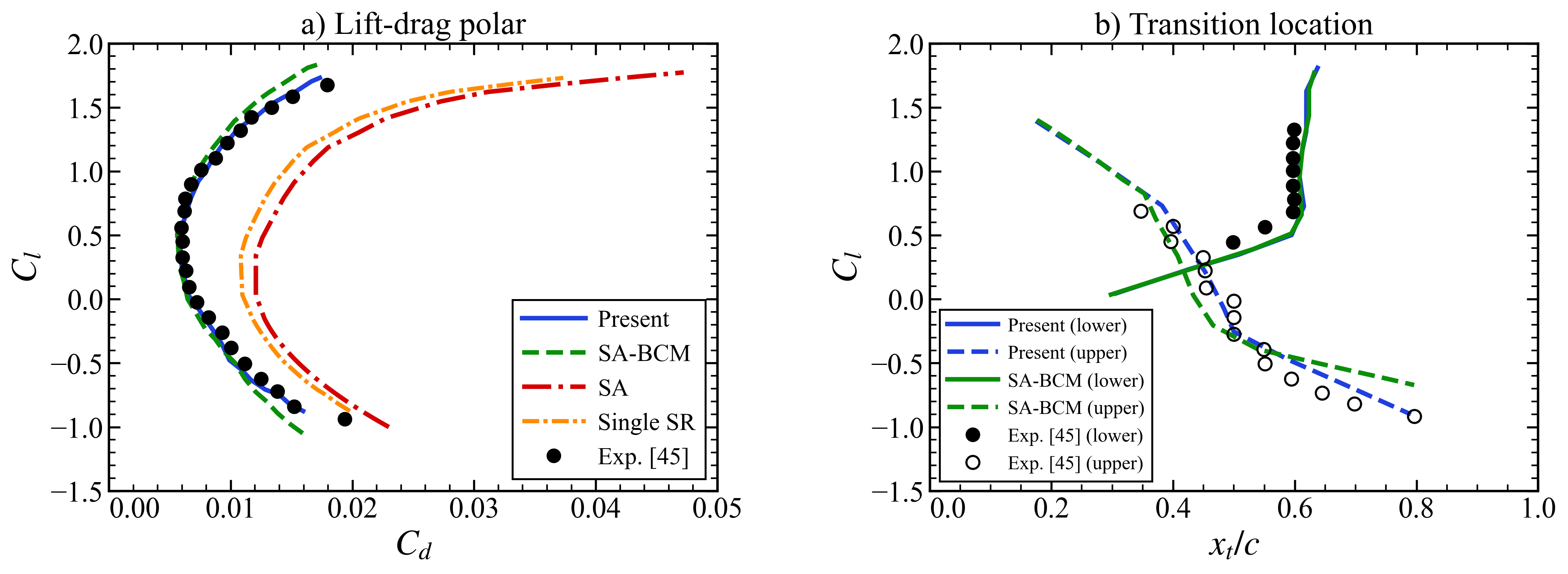}
\caption{Validation of the predicted aerodynamic coefficients and transition locations for the NLF(1)-0416 airfoil under natural-transition conditions ($Re=4\times10^6$, $M=0.1$, $\mathrm{FSTI}=0.15\%$, over a range of $\alpha$).}
\label{fig:nlf_validation_results}
\end{figure}

The results in Fig.~\ref{fig:nlf_validation_results} show that the proposed model predicts the transition-onset location on both the suction and pressure surfaces with reasonable agreement over the investigated range of angles of attack. The SA-based Bas--Cakmakcioglu (SA-BCM) transition model~\cite{cakmakcioglu2020revised,mura2020revised}, used here as an additional SA-based transition-model reference, also predicts the delayed-transition behavior of the NLF(1)-0416 airfoil with a level of accuracy comparable to that of the present model. This agreement indicates that the natural-transition branch retains reasonable predictive performance over angles of attack beyond the single NLF(1)-0416 training incidence.

In contrast, the compact single mixed-mechanism SR diagnostic (Compact Single SR; labeled Single SR in the figures) predicts a substantially reduced laminar region and fails to produce a distinct transition front over much of the investigated angle-of-attack range. This behavior is consistent with the mixed-mechanism diagnostic in Appendix~\ref{app:single_correction}: when natural-, crossflow-, and separation-induced-transition samples are fitted by one compact expression, the resulting model tends to collapse toward an averaged or nearly neutral correction rather than preserving the production suppression required in attached pre-transition regions.

The natural branch also accounts for FSTI through the critical Reynolds number $Re_{\theta c}$ in Eq.~\eqref{eq:re_thetac}. Specifically, $Re_{\theta c}$ is evaluated using the reconstructed local turbulence intensity $Tu_L$, which is bounded below by the prescribed free-stream value $Tu$. This formulation enables the predicted transition onset to shift upstream as FSTI increases within the range considered here. However, it should not be interpreted as an explicit model of bypass-transition physics.

\subsection{Crossflow Transition: 6:1 Prolate Spheroid}

The crossflow-transition capability of the model is validated on the 6:1 prolate spheroid, a widely used three-dimensional test case for crossflow-dominated transition. The spheroid geometry induces a strongly three-dimensional boundary layer with significant crossflow-velocity components, particularly at high angles of attack. Simulations are performed at $Re=6.5\times10^6$, $M=0.13$, and $\mathrm{FSTI}=0.15\%$ for angles of attack $\alpha=10^\circ$ and $20^\circ$. The results are compared against the experimental data of Kreplin \textit{et al.}~\cite{kreplin1985wall}. 

In the converged present-model solution, the near-wall effective production multiplier $\beta_{\mathrm{eff}}$ remains near zero over the windward laminar region, thereby suppressing SA production, and increases across the transition front; as the angle of attack increases from $10^\circ$ to $20^\circ$, the transition front shifts upstream, consistent with enhanced crossflow-instability activity.

\begin{figure}[hbt!]
\centering
\includegraphics[width=1.0\textwidth]{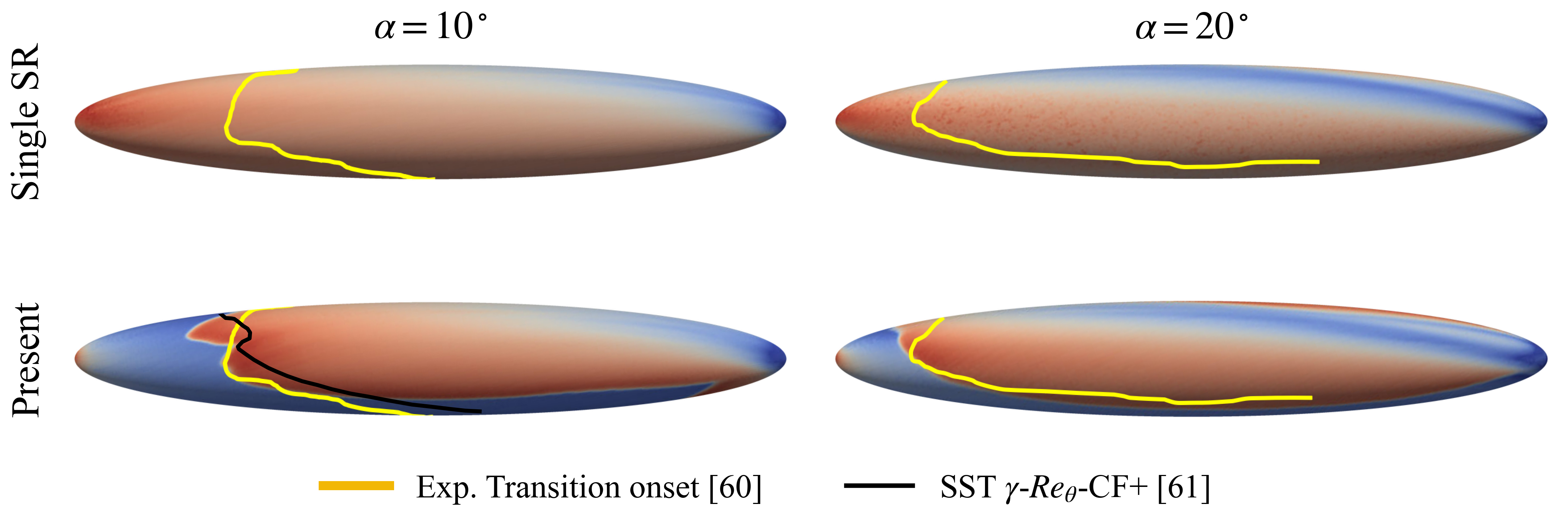}
\caption{Predicted and measured transition-front locations on the 6:1 prolate spheroid for the crossflow-transition validation case ($Re=6.5\times10^6$, $M=0.13$, $\mathrm{FSTI}=0.15\%$) at $\alpha=10^\circ$ and $20^\circ$.}
\label{fig:prolate_transition_front}
\end{figure}

Figure~\ref{fig:prolate_transition_front} compares the transition front predicted by the present model with the experimental transition markers and the published transition front computed by Park \textit{et al.}~\cite{Park2022} using the crossflow-augmented SST $\gamma$--$Re_{\theta t}$--$\mathrm{CF}^+$ model. At $\alpha=10^\circ$, the present model captures the circumferential variation of the transition-onset location; the predicted contour is broadly consistent with the $\gamma$--$Re_{\theta t}$--$\mathrm{CF}^+$ result and shows comparable agreement with the experimental transition markers, without requiring additional transport equations. At $\alpha=20^\circ$, the predicted transition front shifts upstream in accordance with the experimental trend.

In contrast, the baseline SA model and the Compact Single SR diagnostic predict nearly fully turbulent flow and do not recover a distinct crossflow-transition front. Together with the Pareto-front plateau reported in Appendix~\ref{app:single_correction}, this behavior suggests that the crossflow-specific response is weakened when a single compact expression is required to represent natural, crossflow, and separation-induced transition simultaneously. These results support the use of a dedicated crossflow branch and gate.

\begin{figure}[hbt!]
\centering
\includegraphics[width=0.95\textwidth]{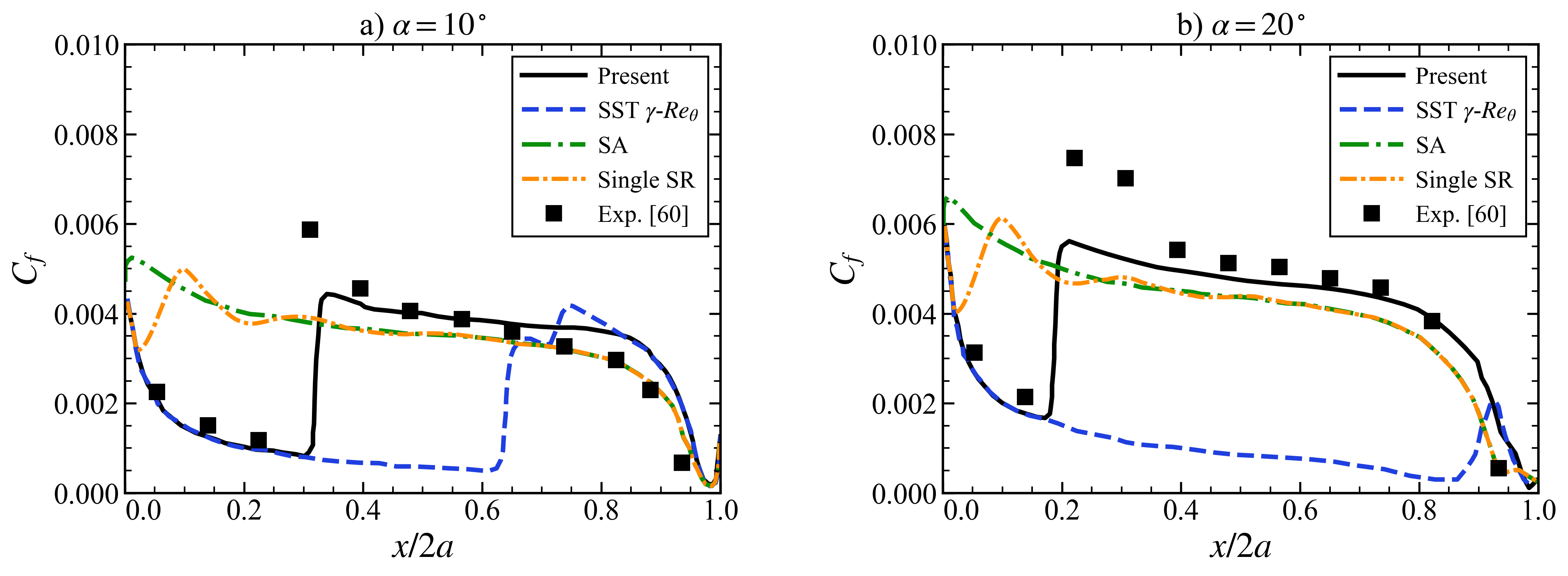}
\caption{Predicted and measured skin-friction coefficient distributions on the 6:1 prolate spheroid ($Re=6.5\times10^6$, $M=0.13$, $\mathrm{FSTI}=0.15\%$) at $\alpha=10^\circ$ and $20^\circ$.}
\label{fig:prolate_cf_distribution}
\end{figure}

The skin-friction distributions in Fig.~\ref{fig:prolate_cf_distribution} show that the present model predicts the transition-associated rise in $C_f$ near the measured streamwise location. The remaining discrepancies near the leeward attachment-line region at $\alpha=20^\circ$ may reflect physical effects not represented by the present criterion, including attachment-line transition. Overall, the results indicate that the local vorticity-direction-twist measure $\Psi$, incorporated through the crossflow criterion $\mathrm{TC1}$, captures the primary crossflow-transition trends in strongly three-dimensional flows without requiring nonlocal $N$-factor integration.

\subsection{Separation-Induced Transition: NACA0015 and SD7003 Airfoils}

The separation-induced-transition branch is evaluated using the NACA0015 airfoil at $\alpha=4^\circ$ and $Re=2\times10^5$ and the SD7003 airfoil at $\alpha=6^\circ$ and $Re=6\times10^4$, both at $\mathrm{FSTI}=0.1\%$, with reference LES data from Ahmed \textit{et al.}~\cite{ahmed2025data}. Both cases exhibit upper-surface laminar separation bubbles. The $\beta_{\mathrm{SIT}}$ branch was calibrated using the NACA0012 airfoil at $\alpha=4^\circ$ ($Re=2\times10^5$) and the SD7003 airfoil at $\alpha=4^\circ$ ($Re=6\times10^4$). Therefore, the NACA0015 case tests transfer to a different airfoil geometry at the same Reynolds number, while the SD7003 case at $\alpha=6^\circ$ evaluates extrapolation to a higher angle of attack at the same Reynolds number as its training condition.

\begin{figure}[hbt!]
\centering
\IfFileExists{Figure/validation/SIT_gate_beta_map.png}{%
\includegraphics[width=0.95\textwidth]{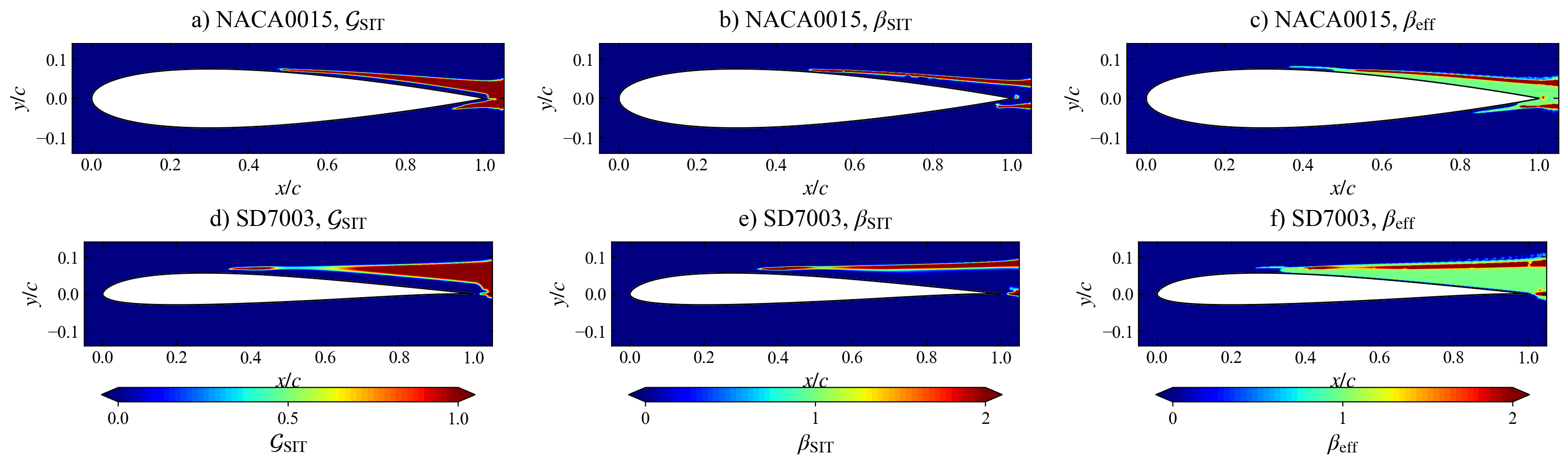}%
}{%
\fbox{\parbox{0.88\textwidth}{\centering Placeholder for NACA0015 and SD7003 maps of $\mathcal{G}_{\mathrm{SIT}}$, $\beta_{\mathrm{SIT}}$, and $\beta_{\mathrm{eff}}$. Replace with gate-activation contours before submission.}}%
}
\caption{Separation-induced-transition onset gate $\mathcal{G}_{\mathrm{SIT}}$, branch correction $\beta_{\mathrm{SIT}}$, and effective correction $\beta_{\mathrm{eff}}$ from the present model for the NACA0015 ($Re=2\times10^5$, $\alpha=4^\circ$) and SD7003 ($Re=6\times10^4$, $\alpha=6^\circ$) separation-induced-transition validation cases, at low Mach number. The $\beta_{\mathrm{SIT}}$ and $\beta_{\mathrm{eff}}$ color scales are clipped at $2$ for display; the underlying branch admits larger values in the separation bubble.}
\label{fig:sit_gate_beta_map}
\end{figure}

To verify the localized activation of the separation-induced correction, Figure~\ref{fig:sit_gate_beta_map} shows the onset gate $\mathcal{G}_{\mathrm{SIT}}$, the separation-induced branch $\beta_{\mathrm{SIT}}$, and the effective correction $\beta_{\mathrm{eff}}$ for both validation airfoils, evaluated from the converged solutions using Eqs.~\eqref{eq:gate_sit}--\eqref{eq:beta_sit}. The onset gate becomes active in the adverse-pressure-gradient region downstream of the separation point, whereas the amplification $\beta_{\mathrm{SIT}}>1$ is confined to a thin near-wall band at the separation-bubble transition location; the effective correction $\beta_{\mathrm{eff}}$ then recovers the baseline production in the turbulent reattached region. The separation-induced contribution is therefore confined to the separated shear layer and reattachment region and remains inactive in the attached laminar boundary layer.

For the NACA0015 airfoil at $\alpha=4^\circ$, the present model captures the main features of the laminar separation bubble, including the separation point, bubble extent, and reattachment location. The $C_f$ distribution exhibits the characteristic near-zero or negative values in the separated region, followed by recovery near reattachment, in agreement with the LES trend. The SST Langtry--Menter model also captures the laminar separation bubble, but slightly overpredicts its length. Conversely, the baseline SA model and the Compact Single SR diagnostic do not reproduce the principal separation-bubble characteristics, instead promoting premature turbulent reattachment near the leading edge; as in the mixed-mechanism diagnostic of Appendix~\ref{app:single_correction}, a single compact expression weakens the production amplification ($\beta>1$) required in separated shear layers when it must simultaneously provide attached-flow suppression ($\beta<1$).

For the SD7003 airfoil at $\alpha=6^\circ$, the separation bubble is larger and the transition process is more complex due to the increased adverse pressure gradient. The present model predicts the elongated separated region and subsequent turbulent reattachment observed in the LES reference. The $C_p$ distribution predicted by the present model captures the suction peak, the pressure plateau over the separation bubble, and the main pressure-recovery trend, although local differences remain near reattachment where the LES shows a sharper recovery. The SST Langtry--Menter model predicts both separation and reattachment farther downstream than the LES reference and the present model. Because the delay in reattachment is more pronounced than the delay in separation, the SST Langtry--Menter model produces an overly long separated region and a delayed pressure recovery.

These results indicate that the $\beta_{\mathrm{SIT}}$ correction, together with the gate function $\mathcal{G}_{\mathrm{SIT}}$, can identify the main separation-dominated regions on both airfoils and apply production amplification within the separated shear layer. While the model captures the principal separation-bubble characteristics, the peak $C_f$ recovery after reattachment is slightly underpredicted. This discrepancy reflects the limitations of a local SA-production correction in reproducing the detailed post-reattachment skin-friction recovery. Nevertheless, the local algebraic formulation predicts transition and reattachment without requiring nonlocal integral boundary-layer parameters or user-specified separation locations.

\begin{figure}[hbt!]
\centering
\includegraphics[width=0.95\textwidth]{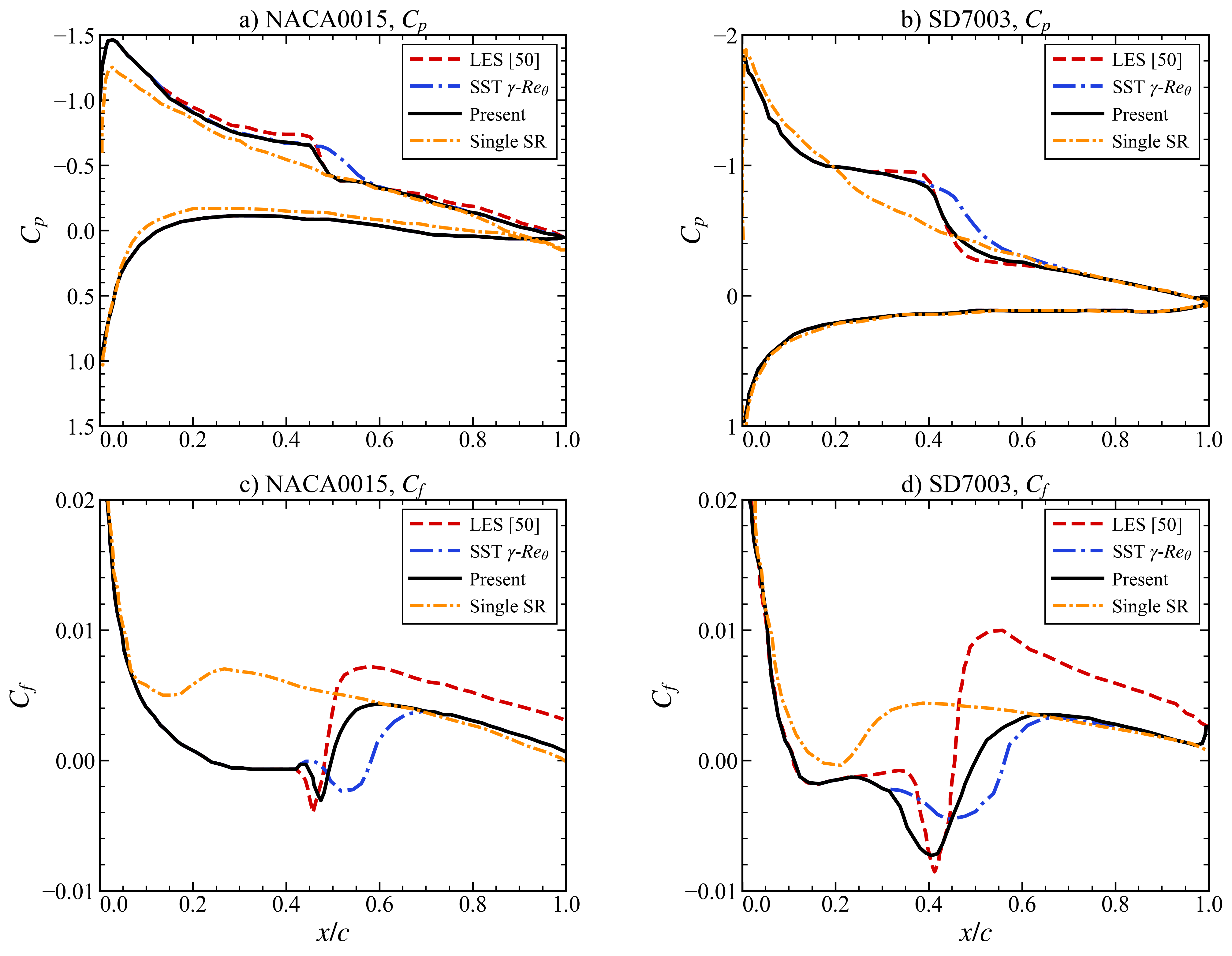}
\caption{Pressure and skin-friction coefficient distributions for the separation-induced-transition validation cases: NACA0015 ($Re=2\times10^5$, $\alpha=4^\circ$) and SD7003 ($Re=6\times10^4$, $\alpha=6^\circ$), at $M=0.1$ and $\mathrm{FSTI}=0.1\%$. Each panel is labeled by airfoil and quantity.}
\label{fig:sit_cp_cf_validation}
\end{figure}

\section{Application to Complex Three-Dimensional Configurations}
\label{sec:complex_3d_applications}

Whereas the canonical validation cases in Section~\ref{sec:model_validation} are designed to emphasize one dominant transition mechanism, this section considers complex configurations in which multiple transition mechanisms may be locally relevant. The CRM-NLF aircraft wing configuration is used as the fixed-wing transonic swept-wing assessment case, where natural-transition and crossflow-sensitive mechanisms can coexist over different portions of the wing. The PSP rotor in hover is used as the rotating-blade assessment case, where natural-transition behavior is expected in low-disturbance attached-flow regions and crossflow-sensitive transition may become relevant in spanwise regions affected by sweep and rotation-induced three-dimensionality.

All complex-configuration computations in this section are performed with KFLOW~\cite{kim2009euler,kflow2}, a compressible RANS solver independent of the DAFoam environment used for field inversion and canonical validation. KFLOW supports structured multi-block and overset-grid topologies, rotating reference frames, and steady multiple-reference-frame calculations. The governing equations are the three-dimensional compressible RANS equations closed by the corresponding SA-based transition correction, baseline SA model, or SST-based transition model. Convective fluxes are discretized using the AUSMPW+ upwind scheme~\cite{kim2001methods} with fifth-order spatial reconstruction through the eMLP-VC limiter~\cite{hong2022enhanced}, whereas viscous fluxes are computed using second-order central differencing. The steady computations are advanced using an implicit second-order backward-difference formulation together with the diagonalized alternating-direction implicit (DADI) method~\cite{PULLIAM1981347}.

This section therefore assesses not only transition-prediction behavior, but also implementation portability of the closed-form correction when it is moved from the DAFoam training environment to an independent RANS solver. The assessment focuses on the agreement of the predicted transition front with available experimental measurements and on computational cost relative to baseline SA and a transport-equation-based transition model at matched convergence.

\subsection{CRM-NLF aircraft wing configuration}

The Common Research Model with Natural Laminar Flow (CRM-NLF) is a NASA research configuration designed to maintain extended regions of natural laminar flow under representative transonic cruise conditions. The CRM-NLF wing is used here to assess the proposed model in a transonic swept-wing flow at $M=0.85$ and $Re=15\times10^6$ based on the mean aerodynamic chord (MAC). The wing-only configuration provides a demanding assessment case because spanwise variations in pressure gradient and sweep-induced three-dimensionality can strongly influence the transition front.

The CRM-NLF aircraft-wing computation uses a structured grid system of approximately $1.1\times10^7$ cells for the wing-only configuration with a symmetry-plane condition; the far-field boundary is placed 50 chord lengths from the surface, the near-wall spacing is maintained at $y^{+}\le0.5$, and the free-stream turbulence intensity is set to $0.24\%$. The predicted transition front is compared with available infrared-thermography transition-onset measurements~\cite{crmnlf_ntf}.

\begin{figure}[hbt!]
\centering
\includegraphics[width=1.0\textwidth]{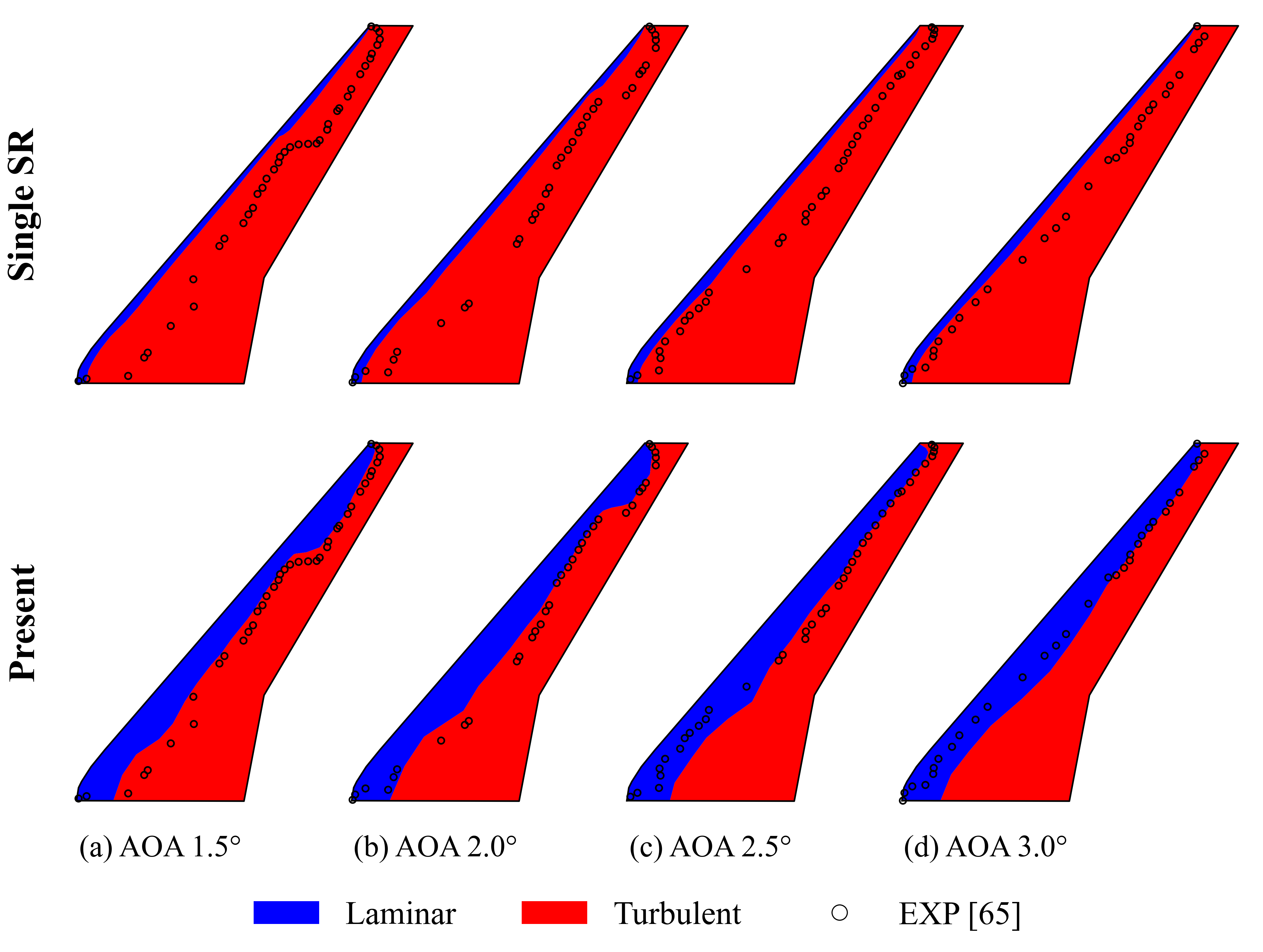}
\caption{Predicted laminar and turbulent surface regions identified using the $i_t=0.95$ criterion on the CRM-NLF wing ($M=0.85$, $Re=15\times10^6$ based on MAC) at different angles of attack.}
\label{fig:CRM_NLF_skin_friction}
\end{figure}

Figure~\ref{fig:CRM_NLF_skin_friction} shows the predicted transition-onset distribution on the CRM-NLF configuration, identified using the turbulence-index criterion. The transition-onset location is taken as the surface location at which Spalart's wall turbulence index reaches $0.95$,
\begin{equation}
i_t=\frac{1}{\kappa u_\tau}\frac{\partial\tilde{\nu}}{\partial n}=\frac{|\nabla\tilde{\nu}|}{0.41\,\sqrt{\nu\,\Omega}},
\label{eq:turb_index}
\end{equation}
and the same threshold is applied consistently to the computed transition fronts in this section. The predicted transition front indicates an extended laminar region over much of the wing. Its spanwise variation is consistent with the combined effects of local pressure-gradient changes and sweep-induced three-dimensionality. 

The compact single mixed-mechanism SR diagnostic (Compact Single SR) predicts a substantially reduced laminar extent and a transition front that is shifted upstream relative to the experimental transition markers and the present mechanism-separated result. Although the Compact Single SR diagnostic does not become strictly turbulent everywhere, it exhibits a baseline-like premature-transition behavior and fails to preserve the extended natural-laminar-flow region on the CRM-NLF wing, consistent with the mixed-mechanism diagnostic of Appendix~\ref{app:single_correction}.

The predicted transition front obtained from the present mechanism-separated model is consistent with the available CRM-NLF transition-onset measurements under cruise conditions and is obtained without case-specific retuning of the branch coefficients, admissibility-gate parameters, or memory-gate parameters. This result supports the use of the mechanism-separated formulation beyond the canonical two-dimensional and swept-wing training configurations.

\subsection{PSP rotor in hover}

The proposed transition model is applied to the PSP (Pressure-Sensitive Paint) rotor, a NASA four-bladed rotor designed for hover-performance testing and transition visualization, to assess the model in a realistic three-dimensional rotating-blade configuration. The PSP rotor is a challenging test case because the local transition behavior, driven by the coexistence of natural and crossflow-sensitive mechanisms, can vary substantially along the blade span. The rotor geometry is obtained from the AIAA Hover Prediction Workshop PSP-rotor database~\cite{aiaahpw_psp}, and the measured hover performance and boundary-layer transition are taken from Overmeyer and Martin~\cite{overmeyer2017measured}. The rotor has four blades, a radius of $1.689$~m, a reference chord of $0.138$~m, and a solidity of $0.1033$; the hover condition considered here corresponds to a rotational speed of $1150$~RPM, i.e., a tip Mach number of $0.58$.

The hover computation is performed on a quarter-domain periodic sector using the azimuthal periodicity of the four-bladed rotor. The overset-grid system for this periodic computational domain consists of a background Cartesian grid containing approximately $1.2 \times 10^7$ cells, with dimensions of $321 \times 195 \times 195$, and one body-fitted structured blade grid containing approximately $8.4 \times 10^6$ cells, with dimensions of $129 \times 257 \times 257$ in the wall-normal, chordwise, and spanwise directions, respectively. The total grid size for the simulated periodic sector is therefore approximately $2.0 \times 10^7$ cells. The blade-surface grid is refined to maintain $y^{+}<0.5$, ensuring adequate near-wall resolution for the SA-based transition correction. The free-stream turbulence intensity is set to $0.09\%$.

The PSP rotor hover computations are conducted at two collective-pitch angles, $\theta_0 = 6^\circ$ and $\theta_0 = 8^\circ$, covering representative blade-loading levels in the range of interest. The figure of merit (FM), which represents the aerodynamic efficiency of the rotor in hover, is adopted as the primary validation metric. It is defined as $\mathrm{FM}=C_T^{3/2}/(\sqrt{2}\,C_Q)$, where $C_T$ and $C_Q$ are the thrust and torque coefficients nondimensionalized by the rotor disk area and tip speed.

\begin{figure}[hbt!]
\centering
\includegraphics[width=0.48\textwidth]{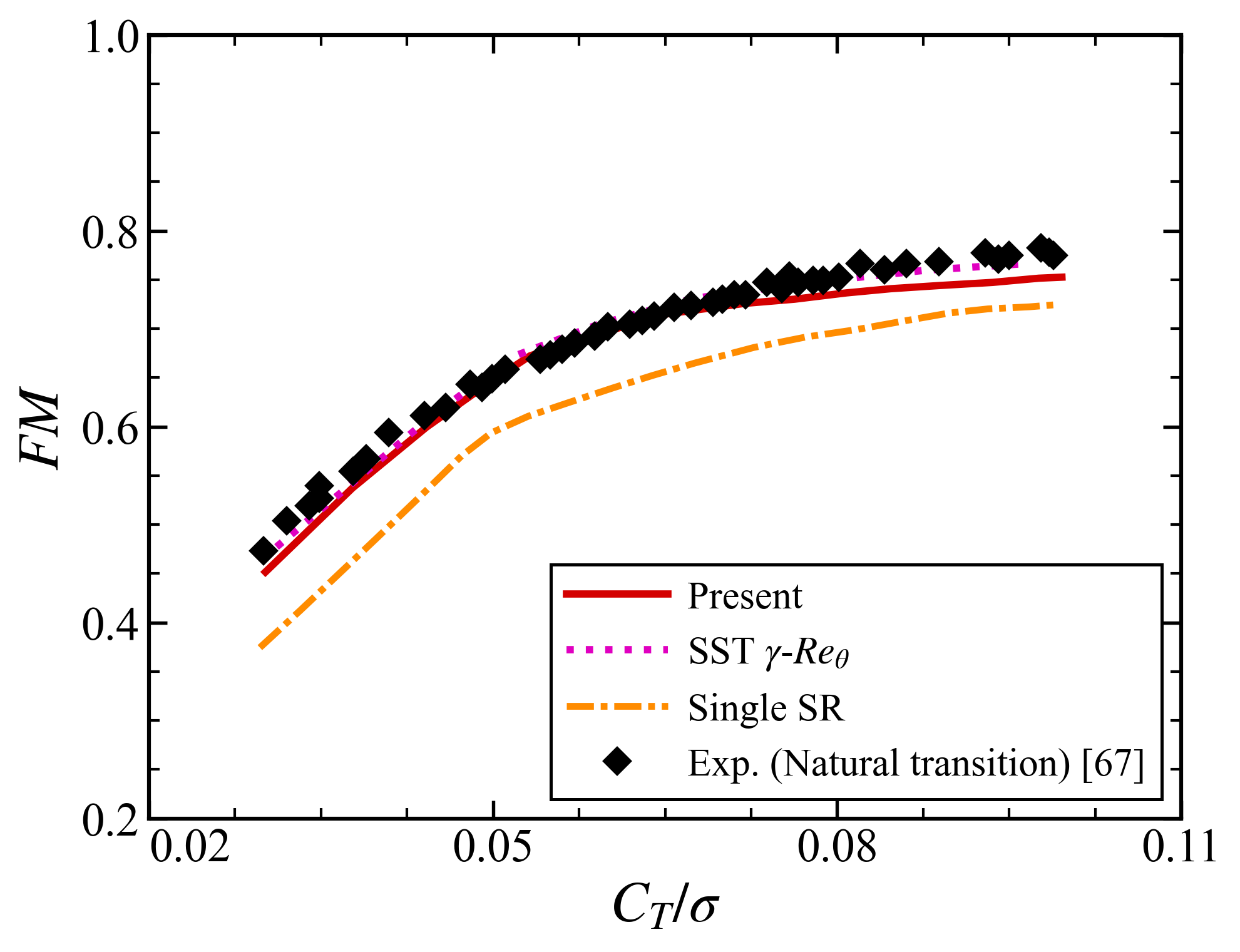}
\caption{Figure-of-merit comparison for the PSP rotor in hover (tip Mach number $0.58$, collective pitch $\theta_0=6^\circ$ and $8^\circ$) as a function of $C_T/\sigma$, comparing the present model, the SST $\gamma$--$Re_{\theta}$ transition model, the Compact Single SR diagnostic, and experimental data.}
\label{fig:psp_rotor_FM}
\end{figure}

\begin{figure}[hbt!]
\centering
\includegraphics[width=1.0\textwidth]{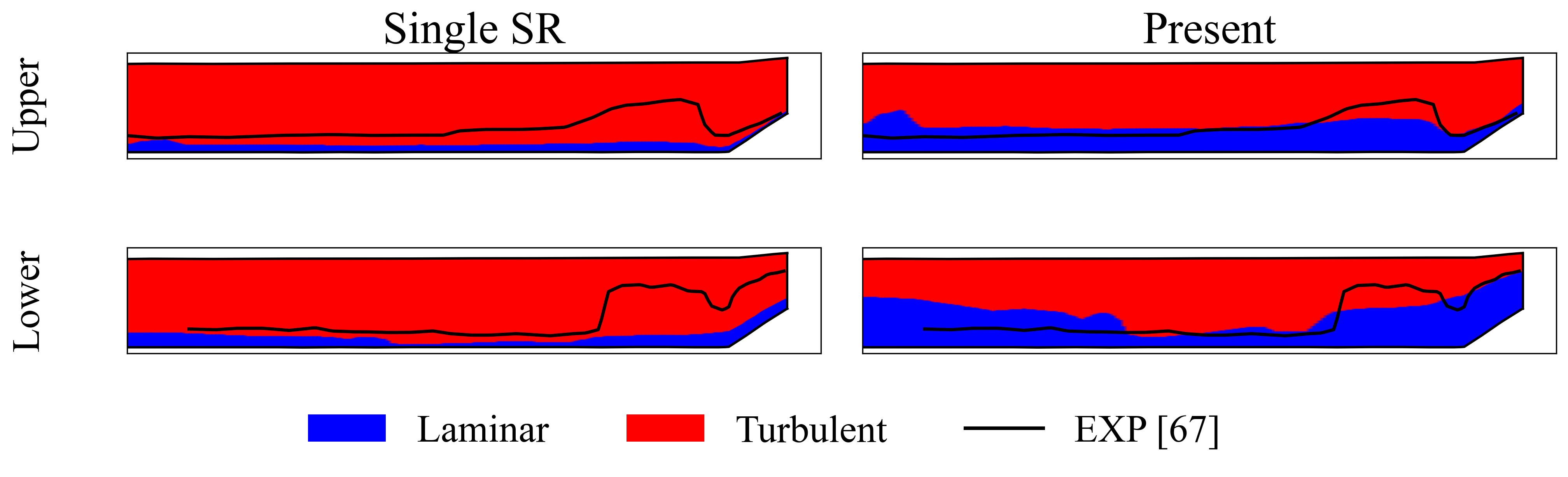}
\caption{Predicted laminar and turbulent surface regions and transition locations on the PSP rotor blade at $\theta_0=6^\circ$ (tip Mach number $0.58$).}
\label{fig:psp_rotor_pitch6_ti_w_CF}
\end{figure}

\begin{figure}[hbt!]
\centering
\includegraphics[width=1.0\textwidth]{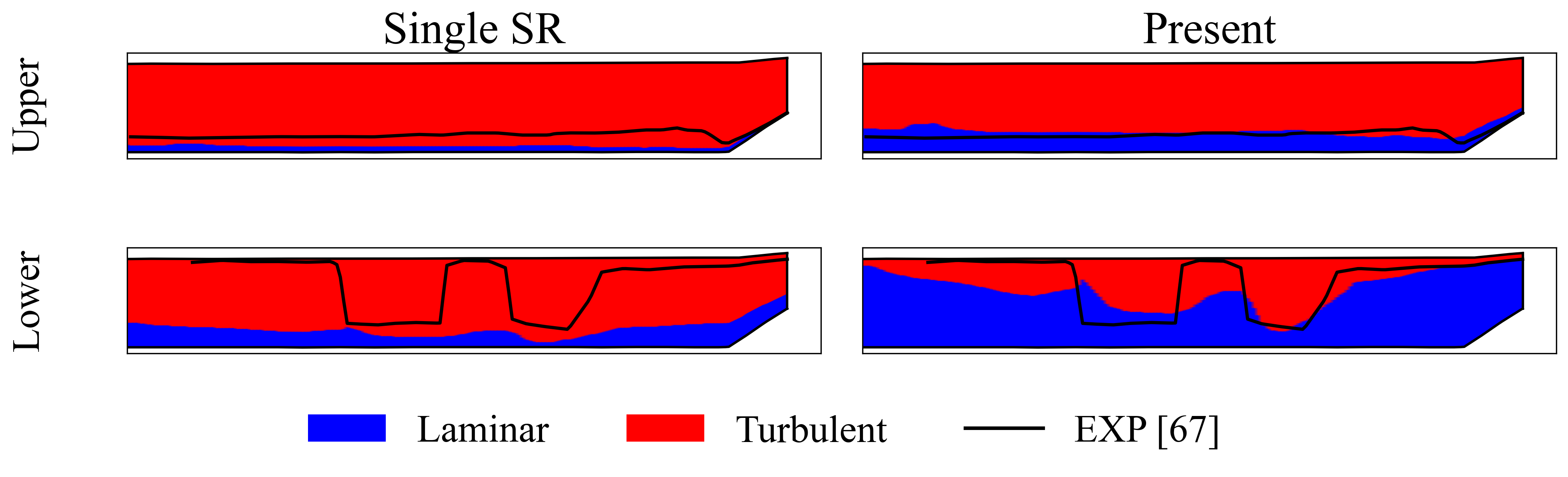}
\caption{Predicted laminar and turbulent surface regions and transition locations on the PSP rotor blade at $\theta_0=8^\circ$ (tip Mach number $0.58$).}
\label{fig:psp_rotor_pitch8_ti_w_CF}
\end{figure}

Figure~\ref{fig:psp_rotor_FM} compares the figure of merit of the PSP rotor in hover predicted by the present model, the SST Langtry--Menter transition model, and the Compact Single SR diagnostic with the experimental data of Overmeyer and Martin~\cite{overmeyer2017measured}. The present model and the SST Langtry--Menter model reproduce the measured $FM$--$C_T/\sigma$ trend reasonably well over the tested operating range. In contrast, the Compact Single SR diagnostic underpredicts the figure of merit over most of the loading range, which is consistent with its premature-transition behavior in the surface transition maps of Figs.~\ref{fig:psp_rotor_pitch6_ti_w_CF} and~\ref{fig:psp_rotor_pitch8_ti_w_CF}. The present algebraic correction provides agreement comparable to the SST Langtry--Menter model while retaining the simpler SA-based algebraic formulation.

The transition-onset distributions at $\theta_0 = 6^\circ$ and $\theta_0 = 8^\circ$ are shown in Figs.~\ref{fig:psp_rotor_pitch6_ti_w_CF} and~\ref{fig:psp_rotor_pitch8_ti_w_CF}. The predicted transition front varies along the blade span, reflecting changes in local Reynolds number, effective sweep, and rotation-induced three-dimensionality. The transition onset is delayed in the lower-loading inboard regions and moves upstream toward the outboard regions, a spanwise movement that is consistent with increasing crossflow sensitivity toward the blade tip.

The Compact Single SR diagnostic does not reproduce this spanwise variation of the transition front. Instead, it predicts a much more upstream and spatially uniform transition pattern, leaving only limited laminar regions compared with the present mechanism-separated model and the PSP-based transition visualization, consistent with the mechanism-conflation trend identified in Appendix~\ref{app:single_correction}. The resulting transition-front pattern from the present model is compared with the available PSP-based transition visualization data without any case-dependent retuning of the model parameters.

\begin{figure}[hbt!]
\centering
\includegraphics[width=0.48\textwidth]{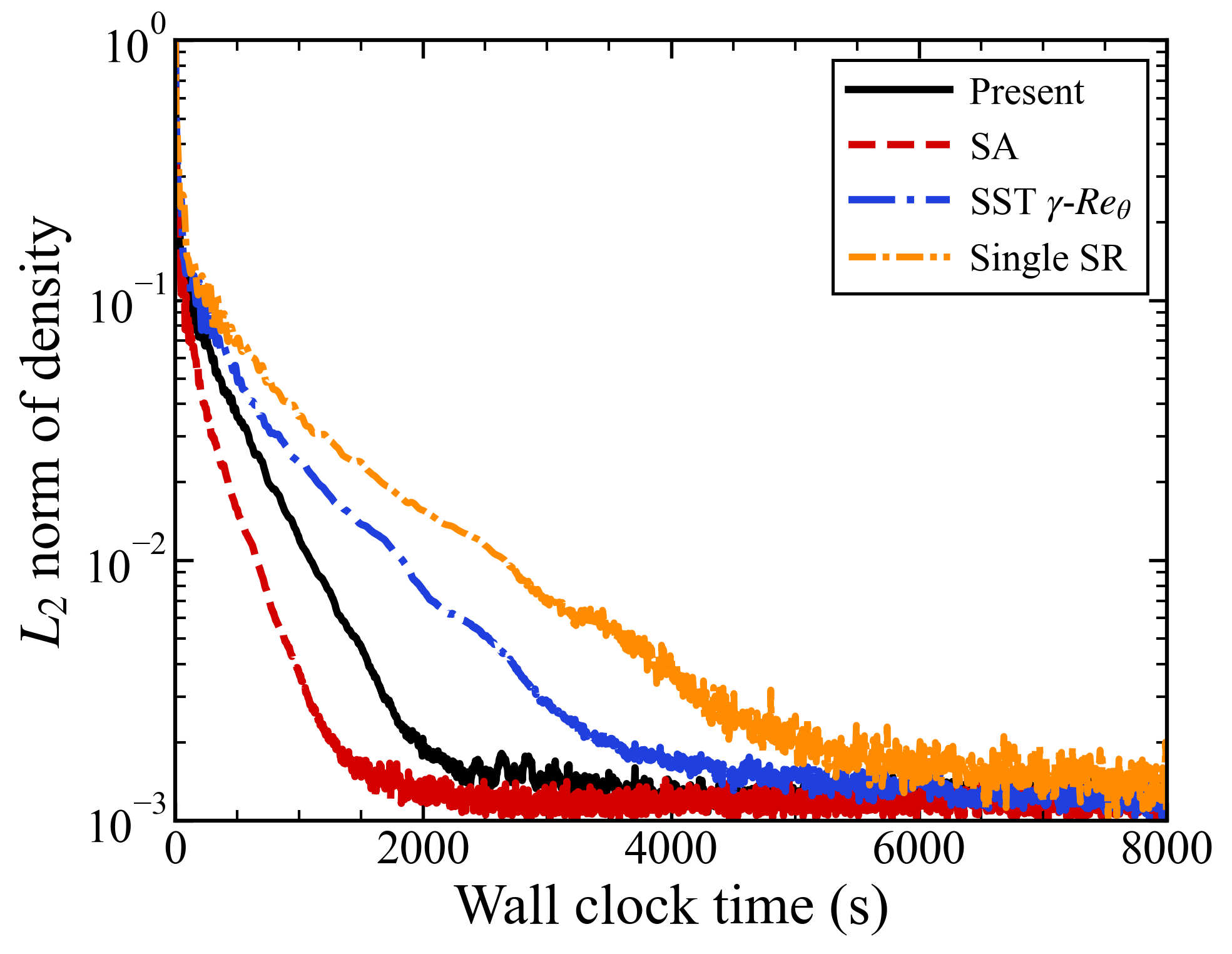}
\caption{Convergence histories of the density residual ($L_2$-norm) as a function of wall-clock time for the PSP rotor hover computation (tip Mach number $0.58$).}
\label{fig:psp_convergence}
\end{figure}

The computational cost of the present algebraic correction is assessed using the PSP rotor hover case. Figure~\ref{fig:psp_convergence} compares the density-residual histories ($L_2$-norm) as a function of wall-clock time. To account for differences in residual-decay rate among the four models, the comparison is based on the wall-clock time required to reach a common residual level of $L_2\approx1.5\times10^{-3}$.

At this residual level, the baseline SA, present, SST Langtry--Menter, and Compact Single SR cases require approximately $1.4\times10^{3}$, $2.3\times10^{3}$, $4.2\times10^{3}$, and $6.0\times10^{3}$~s of wall-clock time, respectively. Thus, the present model incurs an approximately 64\% wall-clock-time overhead relative to the baseline SA model, whereas the SST Langtry--Menter transition model requires approximately 3.0 times the baseline SA cost. Equivalently, the present model reaches the same residual level using approximately 55\% of the wall-clock time required by the SST Langtry--Menter model.

The Compact Single SR curve is shown only as a diagnostic reference and is excluded from the quantitative cost comparison; its poorer transition predictions persist despite continued residual reduction (Appendix~\ref{app:single_correction}). The lower cost of the present model relative to the SST Langtry--Menter model is attributable to the algebraic nature of the transition correction, which does not introduce additional transport equations, although this comparison also reflects the inherent difference between the underlying SA and SST baseline models.

\section{Conclusions}
\label{sec:conclusions}

This paper presented an FISR-assisted, physics-guided development of a mechanism-separated algebraic transition closure within the Spalart--Allmaras (SA) framework: mechanism-specific field inversion supplied the correction targets, symbolic regression identified recurrent local motifs, and the deployed branches consolidated these motifs with literature-derived indicators, admissibility gates, and a turbulence-memory gate. The principal conclusions are summarized as follows.

\begin{enumerate}
    \item Within the operator library, search budget, and implementation-complexity convention used for the diagnostic, mixed-mechanism SR searches in both the reduced and the feature-set-matched full feature spaces plateaued at a balanced pooled MSE of approximately $2.5\times10^{-2}$, whereas the complete separated closure reached $1.0\times10^{-3}$ on the identical pooled data at its implementation complexity of $184$ nodes. This result supports the complete mechanism-separated, physics-guided construction as a compact and interpretable modeling strategy, rather than isolating the effect of branch separation or establishing the impossibility of a global nonlinear mapping.

    \item The consolidated closure consists of explicit algebraic production corrections evaluated from local quantities only; the identical formulation was embedded in the DAFoam development environment and re-implemented in the independent structured-grid solver KFLOW, confirming that no transition transport equations or runtime machine-learning dependencies are required at deployment.

    \item The model was evaluated on canonical and complex three-dimensional configurations and captured the principal transition-front and aerodynamic-performance trends, showing encouraging transfer from the canonical cases to the three-dimensional applications considered. Remaining discrepancies were observed near the prolate-spheroid attachment-line region and in the detailed post-reattachment skin-friction recovery.

    \item For the tested PSP-rotor configuration, hardware, and common density-residual criterion ($L_2\approx1.5\times10^{-3}$), the present implementation required approximately 55\% of the wall-clock time of the SST Langtry--Menter computation, while incurring approximately 64\% overhead relative to the baseline SA model.
\end{enumerate}

The present model remains limited to the transition mechanisms and flow regimes represented in the training and validation data. In particular, phenomena such as bypass transition at elevated free-stream turbulence intensities, roughness-induced transition, shock-induced transition, and leading-edge contamination or icing effects fall outside the scope of the current formulation. Future work should further assess the sensitivity of the gate thresholds, gate parameters, and branch-combination strategy over broader operating conditions and should include additional fixed-wing and rotorcraft configurations to test the generality of the mechanism-separated formulation.

\appendix
\renewcommand{\thesubsection}{\thesection.\arabic{subsection}}
\renewcommand{\thetable}{\thesection\arabic{table}}
\renewcommand{\thefigure}{\thesection\arabic{figure}}
\titleformat{\subsection}
  {\normalsize\bfseries\singlespacing}
  {\thesection.\arabic{subsection}\space}{0pt}{#1}[]
\section{Mixed-Mechanism Symbolic-Regression Diagnostic}
\label{app:single_correction}
\setcounter{table}{0}\setcounter{figure}{0}

\subsection{Mixed-Mechanism Setup}
\label{app:mixed_setup}

To assess whether a single compact algebraic mapping can concurrently represent all three transition mechanisms, a single symbolic-regression correction is trained on the combined field-inversion dataset containing natural-transition, crossflow-transition, and separation-induced-transition samples. This mixed-mechanism diagnostic evaluates the hypothesis that a single compact algebraic mapping can concurrently represent all three transition mechanisms:
\begin{equation}
\beta_{\mathrm{single}}=f_{\mathrm{SR}}(\mathbf{x}),
\label{eq:beta_single_app}
\end{equation}
where $\mathbf{x}$ denotes the selected local-feature vector. The training subset is mechanism balanced and correction balanced, and cells whose post-processed feature reconstruction is inconsistent with the stored solution are excluded: the reconstruction residual is defined as $r=|\beta(\mathbf{x}_{\mathrm{recon}})-\beta(\mathbf{x}_{\mathrm{solver}})|$, and samples with $r>0.1$ are removed, so that the reported Pareto errors measure algebraic expressiveness rather than feature-reconstruction noise. The identical filtered samples are used for every search and comparator; retaining the unfiltered samples raises the common noise floor to approximately $7$--$9\times10^{-2}$ but preserves the model ordering. Detailed exclusion rates and sampling intervals are reported in Supplemental Material~S2.1. Two mixed-mechanism searches were conducted. The reduced search used $\{\lambda_\theta,\,\chi,\,Tu_L,\,Re_V,\,|\Omega|/|S|\}$. The feature-set-matched full search used the untransformed variables $\{Re_V,$ $Re_{\theta c},$ $\lambda_\theta,$ $Tu_L,$ $\chi,$ $\Psi,$ $|\Omega|/|S|\}$, i.e., the full branch-feature union augmented with $|\Omega|/|S|$. The natural-transition raw branch search used the standardized variables listed in Supplemental Material S2, whereas the crossflow and separation branch searches used the untransformed branch-feature union $\{Re_V,$ $Re_{\theta c},$ $\lambda_\theta,$ $Tu_L,$ $\chi,$ $\Psi\}$. Table~\ref{tab:supp_features} summarizes the complete search settings.

\begin{table}[hbt!]
\centering
\captionsetup{labelsep=period}
\caption{Input variables, preprocessing, and sample counts of the symbolic-regression searches.}
\label{tab:supp_features}

\begin{tabularx}{\textwidth}{@{}l>{\raggedright\arraybackslash}X>{\raggedright\arraybackslash}p{3.6cm}r@{}}
\toprule
Search / model & Input variables & Preprocessing & Samples \\ \midrule
Mixed SR (reduced) & $\lambda_\theta,\ \chi,\ Tu_L,\ Re_V,\ |\Omega|/|S|$ & none & 8{,}400 \\
Mixed SR (full-feature) & $Re_V,\ Re_{\theta c},\ \lambda_\theta,\ Tu_L,\ \chi,\ \Psi,\ |\Omega|/|S|$ & none & 8{,}400 \\
Natural raw SR & $Re_\theta{-}Re_{\theta c},\ \nu_t/\nu,\ Tu_L,\ |\Omega|/|S|,\ Re_V$ & z-score (Supplemental Table S1) & 6{,}000 \\
Crossflow raw SR & $Re_V,\ Re_{\theta c},\ \lambda_\theta,\ Tu_L,\ \chi,\ \Psi$ & none & 8{,}400 \\
Separation raw SR & $Re_V,\ Re_{\theta c},\ \lambda_\theta,\ Tu_L,\ \chi,\ \Psi$ & none & 8{,}400 \\
Compact Single SR (comparator) & $\lambda_\theta,\ \nu_t/\nu,\ Tu_L,\ Re_V$ & log/clip $+$ z-score (Sec.~A.3) & 8{,}400 \\
Deployed branches & local physical variables & none & -- \\ \bottomrule
\end{tabularx}
\end{table} The complete mechanism-separated formulation has an implementation complexity of $184$ PySR nodes, or $302$ nodes when repeated subexpressions are counted independently; the mixed-mechanism search was therefore allowed up to $185$ nodes, matching the implementation-level complexity of the proposed formulation, and the per-mechanism branch searches use the corresponding branch-level budget.

\subsection{Pareto-Front Comparison}
\label{app:pareto_comparison}

\begin{figure}[hbt!]
\centering
\includegraphics[width=0.48\textwidth]{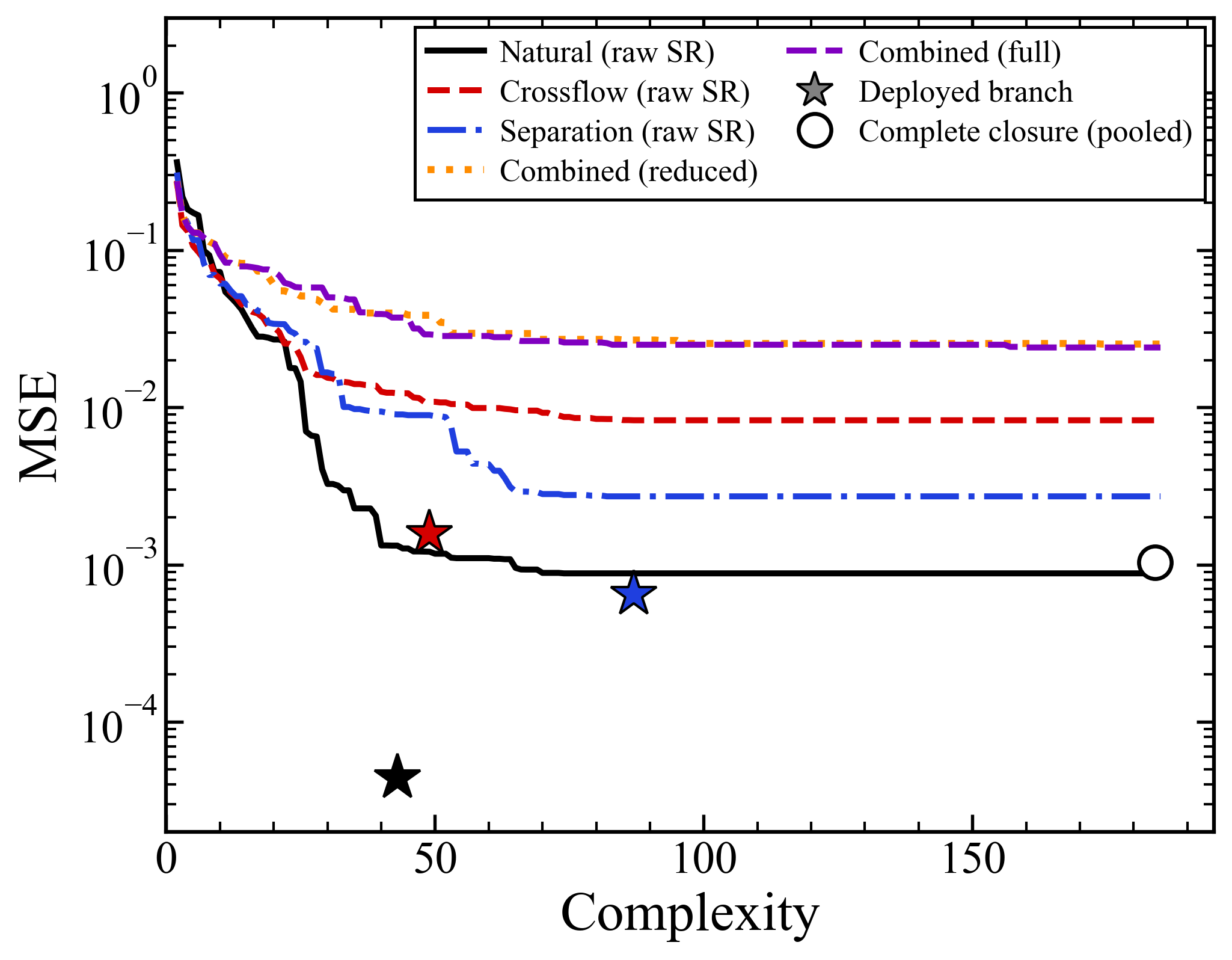}
\caption{Complexity--MSE Pareto fronts of the raw symbolic-regression searches. Stars: deployed consolidated branches on their mechanism-specific subsets; circle: complete separated closure on the pooled subset.}
\label{fig:single_sr_pareto}
\end{figure}

Figure~\ref{fig:single_sr_pareto} shows the resulting complexity--MSE Pareto fronts. Increasing expression complexity does not produce a commensurate reduction in the regression MSE of the single mixed-mechanism expression; the error decreases only weakly beyond a moderate complexity level and saturates near $2.5\times10^{-2}$ even at the matched budget. On their respective mechanism-specific subsets, the deployed consolidated branches reach mechanism-specific diagnostic errors of $4\times10^{-5}$--$1.6\times10^{-3}$ at their implementation node counts ($43$, $49$, and $87$), whereas the representative raw SR expressions retain higher losses, as documented in Supplemental Material S2. A matched pooled comparison, in which the complete separated closure and the best global expressions are evaluated on the identical pooled subset, is reported in Table~\ref{tab:matched_pooled}: the complete closure attains a balanced pooled MSE of $1.0\times10^{-3}$ at implementation complexity $184$, roughly $25$ times below the global-expression floor, and this pooled point is marked separately in Fig.~\ref{fig:single_sr_pareto}. In Fig.~\ref{fig:single_sr_pareto}, the mechanism-specific curves show the best-found losses of the raw branch searches only; the deployed physics-guided consolidated branches are shown as separate star symbols and are not members of the raw fronts. The higher-complexity expressions selected from the single-expression front become increasingly difficult to interpret while still failing to reproduce the heterogeneous correction behavior: under the adopted feature set, operator library, stochastic search protocol, and implementation-complexity convention, increasing the global-expression budget to $185$ nodes did not remove the observed error plateau in any of five independent search seeds; the best losses at the full budget ranged from $2.5\times10^{-2}$ to $4.4\times10^{-2}$, with a median of $3.2\times10^{-2}$. The feature-set-matched full-feature search exhibits the same behavior: across five additional independent seeds with $Re_{\theta c}$, $\Psi$, and $|\Omega|/|S|$ available, the best losses ranged from $2.4\times10^{-2}$ to $3.2\times10^{-2}$ (median $2.6\times10^{-2}$), indicating that the plateau is not an artifact of feature omission.

\begin{table}[hbt!]
\centering
\captionsetup{labelsep=period}
\caption{Matched pooled-task comparison on the identical mechanism-balanced pooled subset (8{,}400 samples): per-mechanism and balanced pooled MSE. Global-SR rows report the best of five seeds.}
\label{tab:matched_pooled}

\begin{tabular}{@{}lccccc@{}}
\toprule
Model & Complexity & Natural & Crossflow & Separation & Balanced pooled \\ \midrule
Global SR, reduced & 185 & $2.3\times10^{-2}$ & $2.5\times10^{-2}$ & $2.7\times10^{-2}$ & $2.5\times10^{-2}$ \\
Global SR, full-feature & 185 & $4.5\times10^{-2}$ & $2.3\times10^{-2}$ & $1.8\times10^{-2}$ & $2.9\times10^{-2}$ \\
Complete separated closure & 184 & $5.1\times10^{-5}$ & $1.5\times10^{-3}$ & $1.6\times10^{-3}$ & $1.0\times10^{-3}$ \\ \bottomrule
\end{tabular}
\end{table}

\subsection{Compact Single SR Comparator}
\label{app:compact_comparator}

The Compact Single SR used in the CFD comparisons is the Pareto-knee expression, of complexity $17$, of a separate mixed-mechanism search conducted on the transformed and standardized four-feature set listed in Table~\ref{tab:supp_features}:
\begin{equation}
\beta_{\mathrm{single}}
=0.6726\left[\tanh\!\left(\hat{z}_{Re_V}\,\hat{z}_{\chi}+e^{\hat{z}_{Tu}}\right)
+\sqrt{\left|-0.2728-\tanh\!\left(-10.087\,\hat{z}_{\lambda_\theta}\right)\right|}\,\right],
\label{eq:single_sr_disclosed}
\end{equation}
where $\hat{z}_{(\cdot)}$ denotes the standardized transformed features
$t_{Re_V}=\log_{10}[\min(Re_V,3000)+1]$, $t_{\chi}=\log_{10}(\nu_t/\nu+1)$, $t_{Tu}=Tu_L$, and $t_{\lambda_\theta}=\lambda_\theta$, standardized as $\hat{z}=(t-\mu)/\sigma$ with
\begin{equation*}
\begin{aligned}
(\mu,\sigma)_{Re_V}&=(2.107,\,0.794),&\quad
(\mu,\sigma)_{\chi}&=(0.558,\,0.804),\\
(\mu,\sigma)_{Tu}&=(4.811,\,6.593),&\quad
(\mu,\sigma)_{\lambda_\theta}&=(-0.0124,\,0.536).
\end{aligned}
\end{equation*}
In the CFD validation cases, this expression does not simultaneously reproduce the production suppression required in attached pre-transition regions and the production amplification required in separated shear layers.

\subsection{Interpretation}
\label{app:interpretation}

The comparison does not establish that every global nonlinear mapping is incapable of representing the pooled correction data. Rather, it shows that, under the selected operator library and solver-deployable complexity budget, the tested compact global symbolic expressions---in both the reduced and the feature-set-matched full feature spaces---exhibit a persistent error plateau and do not reproduce the mechanism-specific correction behavior as accurately as the separated branches. Because the deployed branches also incorporate literature-derived indicators and physics-guided consolidation, this diagnostic is an end-to-end comparison of the complete mechanism-separated modeling strategy with the tested compact global formulations, rather than a controlled architecture-only ablation. The feature-set-matched search equalizes the available local information, although the deployed branches additionally benefit from literature-informed transformations and physics-guided consolidation. The diagnostic therefore supports the use of mechanism separation to obtain a compact, interpretable, and solver-embeddable algebraic closure; it does not establish the general impossibility of a global nonlinear mapping.

\section{Complete Implementation-Ready Model Specification}
\label{app:complete_formulation}
\setcounter{table}{0}\setcounter{figure}{0}

For reference and reproducibility, this appendix collects the complete set of algebraic expressions defining the mechanism-separated correction, using the local variables of Table~\ref{tab:input_features}. The underlying turbulence model is the Spalart--Allmaras model in the SA-noft2 form (the $f_{t2}$ trip term omitted), with the free-stream working variable set to $\tilde{\nu}_\infty = 0.02\,\nu$.

The modified Spalart--Allmaras production term is multiplied pointwise by
\begin{equation}
\beta_{\mathrm{eff}}=\max\left(\beta_{\mathrm{NT}},\,\beta_{\mathrm{CFT}},\,\beta_{\mathrm{SIT}}\right).
\label{eq:beta_eff_complete}
\end{equation}
\subsection{Local Variables and Natural-Transition Threshold}
The natural and crossflow branches use the local onset and turbulence-memory terms
\begin{equation}
Re_V=\frac{d^{2}|\Omega|}{\nu},\qquad
Re_\theta=\frac{Re_V}{2.193},\qquad
\mathcal{T}_1=\frac{\max(Re_\theta-Re_{\theta c},\,0)}{0.002\, Re_{\theta c}},\qquad
\mathcal{T}_2=50\,\frac{\nu_t}{\nu},\qquad
\chi=\frac{\nu_t}{\nu},
\end{equation}
with the pressure-gradient-corrected critical Reynolds number
\begin{equation}
\lambda_\theta=-7.57\times10^{-3}\,\frac{(\mathrm{d}V/\mathrm{d}y)\,d^{2}}{\nu}+0.0128,\qquad
\frac{\mathrm{d}V}{\mathrm{d}y}=\nabla(\hat{\mathbf{n}}\cdot\mathbf{U})\cdot\hat{\mathbf{n}},\qquad
\hat{\mathbf{n}}=\frac{\nabla d}{|\nabla d|},
\end{equation}
\begin{equation}
F_{PG}=
\begin{cases}
\min(1+14.68\,\lambda_\theta,\,1.5), & \lambda_\theta\ge0,\\[2pt]
\min[1-7.34\,\lambda_\theta-7.34\min(\lambda_\theta+0.0681,0),\,3.0], & \lambda_\theta<0,
\end{cases}
\end{equation}
\begin{equation}
Re_{\theta c}=\max\!\left[100+1000\,e^{-Tu_L F_{PG}},\,10^{-3}\right].
\end{equation}
For compressible configurations, this critical Reynolds number is additionally multiplied by a local compressibility factor $f_{cc}$, which was constructed in the present work by fitting the measured Mach-number dependence of the transition Reynolds number from the $10^\circ$-cone in-flight transition experiments of Fisher and Dougherty~\cite{fisher1982inflight}; the factor remains approximately unity in the incompressible limit ($f_{cc}(M_e{=}0)\approx1.008$) and is not applied in the incompressible cases,
\begin{equation}
Re_{\theta c}\;\leftarrow\;f_{cc}\,Re_{\theta c},\qquad
f_{cc}=\sqrt{\,1+\frac{3.3}{1+\exp\!\left[-6.698\left(1.2\,M_e-0.7913\right)\right]}\,},
\end{equation}
where the local isentropic edge Mach number $M_e$ is recovered from the local static pressure $p_{\mathrm{loc}}$ and the free-stream total pressure $p_{0,\infty}$,
\begin{equation}
M_e=\sqrt{\max\!\left[0,\;\frac{2}{\gamma_g-1}\left(\left(\frac{p_{0,\infty}}{p_{\mathrm{loc}}}\right)^{\!(\gamma_g-1)/\gamma_g}\!-1\right)\right]},\qquad
p_{0,\infty}=p_{\infty}\left(1+\frac{\gamma_g-1}{2}M_{\infty}^{2}\right)^{\!\gamma_g/(\gamma_g-1)},
\end{equation}
with $\gamma_g=1.4$ the ratio of specific heats and $M_{\infty}$ the free-stream Mach number; $f_{cc}$ increases $Re_{\theta c}$ with rising local Mach number, delaying transition onset in compressible regions. Here the eddy viscosity is $\nu_t=\tilde{\nu}\,f_{v1}$ from the standard SA relation, and the local turbulence intensity (in percent) is
\begin{equation}
\omega=\frac{|\Omega|}{\sqrt{0.09}},\qquad
k=\nu_t\,\omega,\qquad
Tu_L=\max\!\left[\min\!\left(100\,\frac{\sqrt{2k/3}}{\omega\,d},\;100\right),\;Tu\right].
\end{equation}
\subsection{Crossflow Criterion and Attached-Flow Branches}
The crossflow criterion and its onset term are, with the regularized unit vorticity vector $\mathbf{e}_\omega=\boldsymbol{\omega}/\max(|\boldsymbol{\omega}|,\epsilon_\omega)$ and $\epsilon_\omega=10^{-15}$~s$^{-1}$,
\begin{equation}
\Psi=\left|\hat{\mathbf{n}}\cdot\nabla\mathbf{e}_\omega\right|d,\qquad
\lambda_{cf}=\min\!\left[\max\!\left(-7.57\times10^{-3}\,\frac{(\mathrm{d}V/\mathrm{d}y)\,d^{2}}{\nu}+0.0174,\,0\right),\,0.0477\right],
\end{equation}
\begin{equation}
\begin{aligned}
g_\lambda&=\min\!\left[\max\!\left(\left((27864\,\lambda_{cf}-1962)\lambda_{cf}+54.3\right)\lambda_{cf}+1,\,1\right),\,2.3\right],\\[3pt]
\mathrm{TC1}&=\frac{0.684}{150.8\,g_\lambda}\,\Psi\,Re_V,\qquad
\mathcal{T}_{cf}=\frac{\max(\mathrm{TC1}-1,\,0)}{0.002}.
\end{aligned}
\end{equation}
The shared memory term enters through the smooth gate, giving the attached-flow branches
\begin{equation}
\mathcal{F}_{m}=1-\exp\!\left(-\left[\mathcal{T}_1+\left(\frac{\mathrm{TC1}}{0.45}\right)^{24}+\left(\frac{\chi}{2}\right)^{12}\right]\right),\qquad
\mathcal{T}_2^{\mathrm{eff}}=\mathcal{T}_2\,\mathcal{F}_{m},
\end{equation}
\begin{equation}
\beta_{\mathrm{NT}}=1-\exp\!\left(-\sqrt{\mathcal{T}_1}-\sqrt{\mathcal{T}_2^{\mathrm{eff}}}\right),\qquad
\beta_{\mathrm{CFT}}=1-\exp\!\left(-\sqrt{\mathcal{T}_{cf}}-\sqrt{\mathcal{T}_2^{\mathrm{eff}}}\right).
\label{eq:branch_exponential}
\end{equation}
\subsection{Separation-Induced Branch}
The separation-induced branch is
\begin{equation}
\mathcal{G}_{\mathrm{SIT}}=\min\!\left[\max\!\left(\frac{Re_V/(5\, Re_{\theta c})-1}{0.5},\,0\right),\,1\right],\qquad
F_\chi=\frac{\chi}{5},
\end{equation}
\begin{equation}
F_{\mathrm{reat}}=e^{-(\chi/20)^{4}},\qquad
F_{\mathrm{wake}}=e^{-\min(Re_V/10^{5},\,30)^{2}},\qquad
F_{\mathrm{apg}}=\min\!\left[\max\!\left(\frac{-\lambda_\theta}{0.05},\,0\right),\,1\right],
\end{equation}
\begin{equation}
\beta_{\mathrm{SIT},0}=\min\!\left[2\,\mathcal{G}_{\mathrm{SIT}}\,F_{\mathrm{reat}}\,F_{\mathrm{wake}}\,F_{\mathrm{apg}}\,F_\chi,\;10\right],\qquad
\beta_{\mathrm{pre}}=\max\!\left(\beta_{\mathrm{NT}},\,\beta_{\mathrm{CFT}},\,\beta_{\mathrm{SIT},0}\right),
\end{equation}
\begin{equation}
C_\gamma=\min\!\left[\frac{\max(\beta_{\mathrm{pre}}-0.2,\,0)}{0.8},\,1\right],\qquad
F_{\mathrm{on}}=\min\!\left[\max\!\left(\frac{Re_V}{2420}-1,\,0\right),\,3\right],
\end{equation}
\begin{equation}
\beta_{\mathrm{SIT}}=\beta_{\mathrm{SIT},0}
+\frac{C_\gamma\,F_{\mathrm{on}}\,\max(3\nu-\nu_t,\,0)\,|\Omega|}{\mathrm{Prod.}_{\mathrm{safe}}},\qquad
\mathrm{Prod.}_{\mathrm{safe}}=C_{b1}\tilde{S}\tilde{\nu}+\epsilon_{P},
\label{eq:beta_sit_app}
\end{equation}
where $\mathrm{Prod.}_{\mathrm{safe}}$ is the baseline SA production of the modified SA transport equation of the main manuscript augmented with the additive floor $\epsilon_{P}=10^{-30}$~m$^2$/s$^2$ used in the implementation; the role of the reattachment term is discussed in Section~\ref{sec:branches} and Supplemental Material S3.

\subsection{Provenance and Calibration of Model Components}
\label{app:provenance}

Table~\ref{tab:provenance} states, for every structural element and coefficient of the closure, whether it is inherited from the literature, identified by the mechanism-specific symbolic-regression search, or fitted by the authors to independent data. Only the algebraic Bas--Cakmakcioglu onset and memory scalings ($0.002$ and $50$), the local transition-onset and pressure-gradient correlations, and the Arnal--Menter--Smirnov crossflow criterion are inherited in their cited forms and excluded from the search; all gate thresholds, exponents, limiter scales, and the separation-branch amplitude and bound are identified through FISR-assisted calibration over the ranges in Table~\ref{tab:provenance}. Candidates were ranked by the mechanism-specific correction-balanced mean-squared error; candidates exhibiting solver oscillations, false-positive activation, or a training-case transition-onset error greater than $0.05c$ were rejected; among the remaining candidates, the lowest-complexity candidate within $5\%$ of the minimum loss was selected, without using validation-case data. The full calibration procedure and representative raw expressions are provided in Supplemental Material~S2.2 and~S2.3. In Table~\ref{tab:provenance}, ``regression loss'' denotes the mechanism-specific subset mean-squared error, and ``false-positive exclusion'' and ``bubble-recovery accuracy'' denote, respectively, suppression of spurious separation-branch activation on the natural-transition training cases and reproduction of the post-bubble skin-friction recovery.

\begin{table}[hbt!]
\centering
\captionsetup{labelsep=period}
\caption{Provenance and calibration of model components.}
\label{tab:provenance}

\begin{tabularx}{\textwidth}{@{}>{\raggedright\arraybackslash}p{4.3cm}lcl>{\raggedright\arraybackslash}X@{}}
\toprule
Term or coefficient & Origin & Calibrated & Search range & Selection basis \\ \midrule
$Re_\theta = Re_V/2.193$ & Literature & No & -- & Inherited relation \\
Onset/memory scales $0.002$, $50$ & Literature & No & -- & Inherited coefficients \\
$F_{PG}$, $Re_{\theta c}$ correlation & Literature & No & -- & Inherited correlation \\
$\lambda_{cf}$, $g_\lambda$, $0.684/150.8$ & Literature & No & -- & Inherited criterion \\
Memory-gate threshold $0.45$ & FISR-assisted calibration & Yes & $[0.2,\,1.0]$ & Regression loss, Pareto knee \\
Memory-gate exponents $24$, $12$ & FISR-assisted calibration & Yes & powers $2$--$32$ & Sharpest stable gate \\
Separation-gate coefficient $5$ & FISR-assisted calibration & Yes & $[2,\,8]$ & Regression loss \\
Reattachment limiter scale $\chi=20$ & FISR-assisted calibration & Yes & $[5,\,50]$ & Regression loss \\
Wake-exclusion scale $Re_V=10^{5}$ & FISR-assisted calibration & Yes & $[10^{4},\,10^{6}]$ & False-positive exclusion \\
APG threshold $\lambda_\theta=0.05$ & FISR-assisted calibration & Yes & $[0.01,\,0.1]$ & Regression loss \\
Onset scale $Re_V=2420$ & FISR-assisted calibration & Yes & $[10^{3},\,10^{4}]$ & Bubble-recovery accuracy \\
Separation amplitude $2$, bound $10$ & FISR-assisted calibration & Yes & $[1,\,5]$, $[2,\,20]$ & Regression loss \\
$\max$ branch selector & Physics-guided architecture & No & -- & Avoids double counting of branches \\
$f_{cc}$ compressibility factor & Author fit & No & -- & Fit to Ref.~\cite{fisher1982inflight} \\ \bottomrule
\end{tabularx}
\end{table}

\subsection{A Priori Component-Influence Analysis}
\label{app:ablation}

Table~\ref{tab:ablation} quantifies the influence of each model component through a cumulative a priori component-influence analysis evaluated pointwise on the mechanism-balanced pooled correction data described in Supplemental Materials S1 and S2. Because the pooled samples are extracted from converged solutions, each variant is compared against the complete formulation evaluated at the same feature values, which isolates the structural effect of the removed components from feature-reconstruction noise. This analysis quantifies the structural influence of each component and should not be interpreted as an independent a posteriori accuracy validation. Starting from the literature correlations alone, each added component reduces the deviation from the complete model; the component-wise trends, including the strongly coupled effect of the strengthened separation coefficients and their admissibility gates, are discussed in Supplemental Material S3.

\begin{table}[hbt!]
\centering
\captionsetup{labelsep=period}
\caption{Mean-squared change induced by cumulative model components relative to the complete closure at fixed feature values.}
\label{tab:ablation}

\begin{tabular}{@{}lcccc@{}}
\toprule
Variant & Natural & Crossflow & Separation & Balanced \\ \midrule
Literature correlations only (natural branch) & 0.100 & 0.265 & 0.101 & 0.156 \\
$+$ literature crossflow/separation branches, $\max$ selector & 0.081 & 0.186 & 0.083 & 0.117 \\
$+$ mechanism-specific separation coefficients & 0.066 & 0.230 & 0.069 & 0.121 \\
$+$ separation admissibility gates & 0.059 & 0.199 & 0.065 & 0.108 \\
$+$ turbulence-memory gate & 0.002 & 0.091 & 0.003 & 0.032 \\
Complete model ($+$ reattachment acceleration) & 0 & 0 & 0 & 0 \\ \bottomrule
\end{tabular}
\end{table}

\FloatBarrier

\section*{Acknowledgments}
The authors thank Professor Soo Hyung Park for providing the KFLOW source code for the current study.

\bibliography{sample}

\end{document}